\documentclass[fleqn,10pt]{wlscirep}
\usepackage[utf8]{inputenc}
\usepackage[T1]{fontenc}

\newcommand{\red}[1]{{\color{black}#1}} 
\newcommand{\magenta}[1]{{\color{black}#1}} 

\newcommand{\HK}[1]{{\color{black}#1}}
\newcommand{\HKcomment}[1]{{\color{black}#1}}
\newcommand{\pr}[1]{{\color{black}#1}}
\newcommand{\prnc}[1]{{\color{black}#1}}
\newcommand{\prif}[1]{{\color{black}#1}}
\newcommand{\prred}[1]{{\color{black}#1}}

\newcommand{\tabref}[1]{Table.\ \ref{#1}}
\newcommand{\figref}[1]{Fig.\ \ref{#1}}

\newcommand{\Miatm}{M_{\rm i}^{\rm atm} }

\newcommand{\Msil}{M_{\rm sil} }
\newcommand{\Misil}{M_{\rm i}^{\rm sil} }
\newcommand{\Mmet}{M_{\rm met} }
\newcommand{\Mimet}{M_{\rm i}^{\rm met} }

\newcommand{\Xmomet}{X{\rm _{met}^{MO}} }
\newcommand{\Cisil}{C_{\rm i}^{\rm sil} }
\newcommand{\Cimet}{C_{\rm i}^{\rm met} }
\newcommand{\Dims}{D_{\rm i}^{\rm met/sil} }
\newcommand{\Dcms}{D{\rm _C}^{\rm met/sil} }
\newcommand{\Dnms}{D_{\rm N}^{\rm met/sil} }
\newcommand{\Dhms}{D_{\rm H}^{\rm met/sil} }
\newcommand{\miatm}{m_{\rm i}^{\rm atm} }

\newcommand{\Xmet}{x{\rm _{met}} }

\newcommand{\train}{\tau_{\rm rain-out} }
\newcommand{\tacc}{\tau_{\rm accretion} }

\title{Numerous chondritic impactors and oxidized magma ocean set Earth's volatile depletion}

\author[1,*]{Haruka Sakuraba}
\author[2]{Hiroyuki Kurokawa}
\author[2]{Hidenori Genda}
\author[1]{Kenji Ohta}
\affil[1]{Department of Earth and Planetary Sciences, Tokyo Institute of Technology, Ookayama, Meguro-ku, Tokyo, 152-8551, Japan}
\affil[2]{Earth-Life Science Institute, Tokyo Institute of Technology, Ookayama, Meguro-ku, Tokyo, 152-8550, Japan}

\affil[*]{sakuraba@eps.sci.titech.ac.jp}

\begin{abstract}
Earth's surface environment is largely influenced by its budget of major volatile elements: carbon (C), nitrogen (N), and hydrogen (H).
\pr{Alt}hough the volatiles on Earth are thought to \pr{have been} delivered by chondritic materials, the elemental composition of the bulk silicate Earth (BSE) shows depletion in the order of N, C, and H. 
Previous studies \pr{have} concluded that non-chondritic materials are needed for this depletion pattern. 
Here\pr{,} we model the evolution of the volatile abundances in the atmosphere, oceans, crust, mantle, and core through the accretion history by considering elemental partitioning and impact erosion. 
We show that the \pr{BSE} depletion pattern can be reproduced from continuous accretion of chondritic bodies by the partitioning of C into the core \magenta{and H storage in the magma ocean} in the main accretion stage and atmospheric erosion of N in the late accretion stage.
Th\pr{is} scenario requires \HK{a relatively oxidized} magma ocean \red{($\log_{10} f_{\rm O_2}$ $\gtrsim$ \magenta{$\rm IW$}$-2$, where $f_{\rm O_2}$ is the oxygen fugacity\magenta{, ${\rm IW}$ is $\log_{10} f_{\rm O_2}^{\rm IW}$, and $f_{\rm O_2}^{\rm IW}$ is} $f_{\rm O_2}$ \magenta{at} the iron-w\"{u}stite buffer}), the dominance of small impactors in the late accretion, and the storage of H and C in oceanic water and carbonate rocks in the late accretion stage, all of which are naturally expected from the formation of an Earth-sized planet in the habitable zone.
\end{abstract}
\begin{document}

\flushbottom
\maketitle
%
%
\thispagestyle{empty}

\section*{Introduction}

Earth's major volatile elements\pr{---}carbon (C), nitrogen (N), and hydrogen (H)\pr{---}are the main components of the atmosphere and oceans and the key elements for life. 
The budget of th\pr{e}se major volatiles in the bulk silicate Earth (including the atmosphere, oceans, crust, and mantle\red{:} hereafter BSE) influences the volume of the oceans and the atmospheric inventory of C (CO$_2$) and N (N$_2$), and consequently, Earth's habitable environment \cite{abbot2012,wordsworth2013,foley2015}. 
Similar isotopic compositions of volatiles in Earth and chondrites suggests \pr{that} delivery was made chiefly by chondritic materials \cite{marty2012}. In contrast, \pr{al}though their absolute abundances are largely uncertain, the abundances of C-N-H in BSE relative to chondrites are known to have a V-shaped depletion pattern \cite{hirschmann2016,halliday2013}. 
\pr{Owing to this} discrepancy\pr{,} the origin of the major volatile elements on Earth \pr{remains unclear} \HK{\cite{hirschmann2016}}. 

\begin{figure}[ht]
\centering
\includegraphics[width=\linewidth]{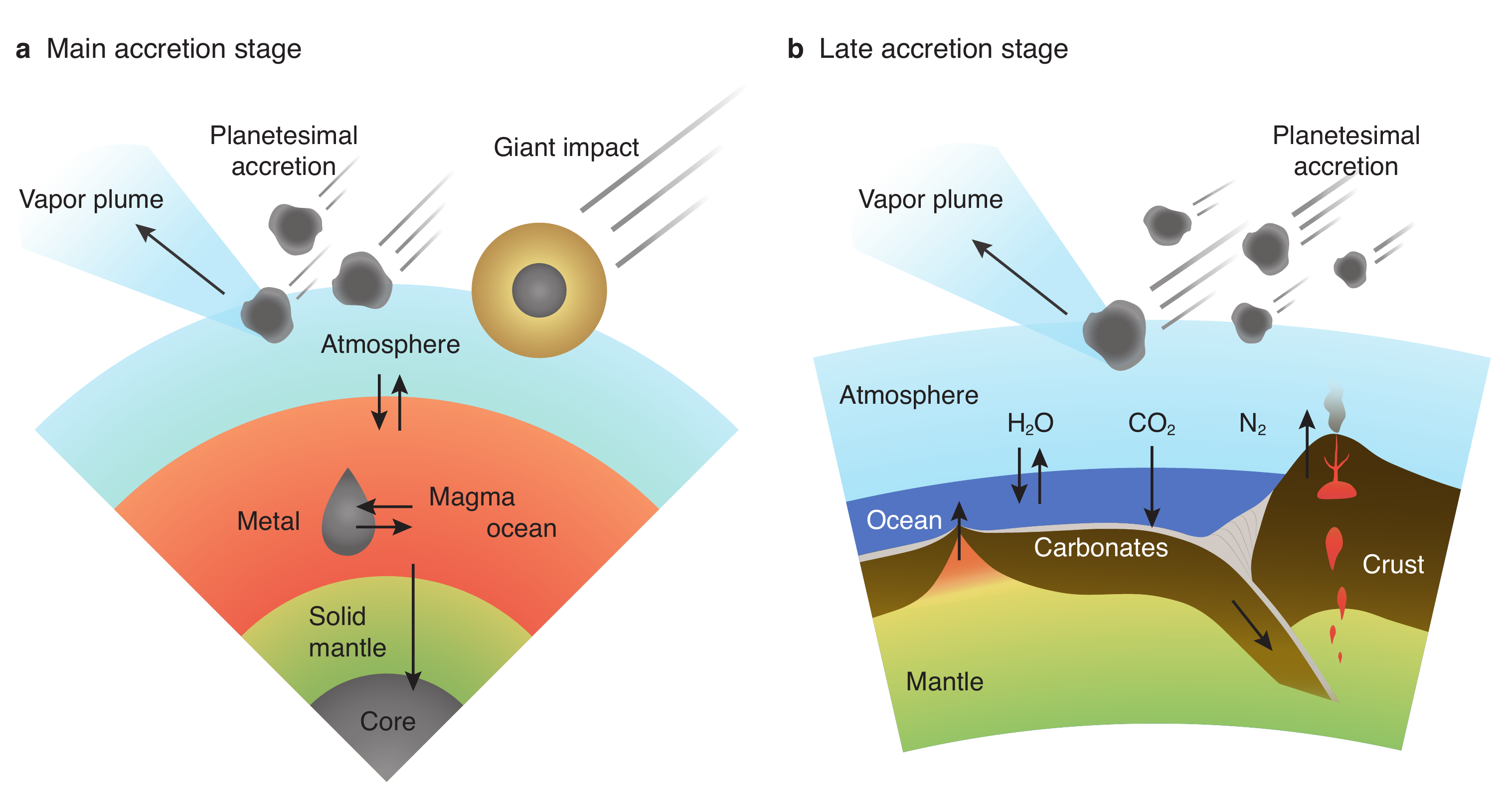}
\caption{{\bf Cartoon of element partitioning processes during Earth's accretion \prif{according to} our model.} Accreting planetesimals and giant impactors deliver volatiles and simultaneously form  a vapour plume eroding the atmosphere. {\bf a:} Model for the main accretion stage (10\pr{\%} to 99.5\% \pr{of} the Earth's mass). Equilibration among the magma ocean (silicate melt), liquid metal droplets transiting to the core, and the overlying atmosphere are achieved according to each metal-silicate partitioning coefficient and solubility. {\bf b:} Model for the late accretion stage after the solidification of the magma ocean (the last 0.5\%). We consider the liquid water oceans and the carbon\red{ate}-silicate cycle to be driven by plate tectonics on the surface. In this stage, most H and C on Earth are stored in the oceans and carbonate rocks, respectively. Numerous impactors can selectively erode N.}
\label{fig:modelimage}
\end{figure}

The composition of major volatiles in the BSE should have been modified by element partitioning processes, including removal by core-forming metal\pr{s} and by atmospheric escape (\figref{fig:modelimage}).
\pr{Alt}hough the origins and timing of volatile delivery are still debated, the delivery \pr{during} the main accretion stage is  expected \pr{based on} the planet formation theory\pr{,} which involves large-scale dynamic evolution, such as the Grand Tack model \cite{walsh2011}. Recent isotopic analyses point to the late-accreting bodies being composed of enstatite chondrites \cite{dauphas2017} or carbonaceous chondrites\cite{fischergodde2020}, suggesting that \pr{volatile delivery} continued \pr{in}to the late stage. 
On the growing proto-Earth with planetesimal accretion and several giant impacts, the formation of magma oceans allow\pr{ed} volatiles to be stored within \HK{the} \red{magma ocean} \cite{elkins2008}. 
Core-forming metal could have removed \pr{some} of the iron-loving elements (siderophiles) from the magma ocean during the main accretion stage \cite{dasgupta2019}. 
\pr{V}olatiles partitioned into the atmosphere (atmophiles) \pr{were} continuously removed via atmospheric erosion caused both by small planetesimal accretion\cite{deNiem2012} and giant impacts\cite{gendaabe2005}.
The successive late accretion after the solidification of the magma ocean further remove\pr{d} and replenishe\pr{d} volatile elements\cite{sakuraba2019}.

\HK{\pr{Al}hough several previous studies \pr{have attempted to} explain the depletion pattern\pr{s} of major volatile elements in BSE \cite{bergin2015,hirschmann2016,grewal2019}, the} evolution of the volatile composition through the full accretion history has not been simulated.
\HK{The} previous studies employed ad hoc models where a single-stage metal-silicate equilibration event and complete/negligible atmospheric loss \pr{were} assumed. 
Hirschmann\cite{hirschmann2016} showed that the combination of core segregation and atmospheric blow off would leave BSE with low C/H \pr{and} C/N ratios \pr{compared with} accreted material, and he concluded that \pr{the} BSE's high C/N ratio requires late accreting bodies with elevated C/N ratios compared \pr{with} chondrites. 
Other works \pr{have} attributed the discrepancy largely to the accretion of thermally processed or differentiated, non-chondritic bodies\cite{bergin2015,grewal2019}, which are hypothetical and may not satisfy the isotopic constraints\cite{dauphas2017}. 
\HK{\pr{Here,} we consider another mechanism \pr{that} can fractionate C/N:} the preferential loss of N relative to C and H by impacts during the late accretion \HK{stage, where N is partitioned into the atmosphere\pr{,} while C and H \pr{are partitioned} into the oceans and carbonate rocks}\cite{sakuraba2019}. \HK{\pr{This work builds on previous studies}\cite{bergin2015,hirschmann2016,grewal2019} \prif{in terms of} volatile element partitioning\pr{, but makes} improvements to simulate core formation and atmospheric loss as continuous processes rather than single stage events.} \par

In this \pr{study}, we aim\prif{ed} to reproduce the V-shaped C-N-H pattern by considering realistic processes to the extent of today’s observational uncertainties. 
We modelled the evolution of the volatile abundances in the atmosphere, oceans, crust, mantle, and core through the full accretion by taking elemental partitioning and impact erosion into account. \red{\figref{fig:modelimage} shows \pr{a} schematic image of our model setting.} The main and late accretion stages were modelled separately, and the masses of C, N, and H in each reservoir were computed using a multiple-boxes model (Methods). \HK{We assumed the existence of the oceans and the active \HK{carbonate}-silicate cycle in the late accretion stage\prif{; the} validity \prif{of this assumption is discussed.}} 
We explored the plausible accretion scenarios \pr{that} reproduce the current BSE's C-N-H composition pattern from the accretion of chondritic bodies. The major parameters \pr{were} the size distribution of planetesimals in each stage\pr{;} the number of giant impacts\pr{;} the redox state of \pr{the} magma ocean\pr{,} which controls the solubility and metal-silicate partitioning coefficient of volatiles\pr{;} the composition of impactors\pr{;} and the total mass of the late accretion. Our nominal model\prred{, which is a successful case,} assumes \pr{a} volatile supply from CI chondrite-like building blocks, \pr{an} oxidized magma ocean \HK{($\log_{10} f_{\rm O_2}$ $\sim$\magenta{\rm IW+1}}\magenta{, where $f_{\rm O_2}$ is the oxygen fugacity\magenta{, ${\rm IW}$ is defined hereafter as $\log_{10} f_{\rm O_2}^{\rm IW}$, and $f_{\rm O_2}^{\rm IW}$ is} $f_{\rm O_2}$ \magenta{at} the iron-w\"{u}stite buffer)}, \pr{a} single giant impact, and \pr{a} change in planetesimal size distribution with time, and 0.5 wt.\% late accretion. \prred{\figref{fig:patternevolution} and \figref{fig:CNHin3layer} \magenta{show} the evolution of major volatile abundances for this successful case.} \red{As the composition of building blocks, a mixture of CI chondrite-like impactors (12 wt.\%) and dry objects (88 wt.\%) \pr{was} fixed by exploring the best fit homogeneous accretion (see Supplementary Information). In order to understand the physical behaviours of the volatile element partitioning, we also calculated the \HK{evolution} for other cases \pr{with different} impactor size distribution\pr{s}, accretion scenarios, amount\pr{s} of late accretion\red{, and redox states of the magma ocean} (\figref{fig:parametersv}). In \pr{the} Supplementary Information, we \HK{show the results for the cases where we assume different source for volatile elements (enstatite chondrites) and the range of} partitioning \HK{coefficients and solubilities.} \HK{We} confirmed that o}ther parameters \red{such as the magma ocean depth, m\pr{e}tal/silicate ratio, surface temperature during magma ocean stage, \pr{and} efficiency of impact erosion by \HK{a} giant impact ha\pr{d}} \pr{only} minor effects \HK{(Fig. \prred{S1})}. 
\magenta{The uncertainties \HKcomment{in the final volatile abundances in BSE caused by these} minor parameters, except for the magma ocean depth, are smaller than 10\%. \HKcomment{As to} the magma ocean depth, the uncertainties differ by species and the redox \HKcomment{state} of the magma ocean: a factor of $\sim$2 for H in the oxidized model, $\sim$40\% for N and $\sim$15\% for H in the reduced model, and smaller than 10\% for \HKcomment{the} others \HKcomment{(Fig. S1 for the oxidized model and the figure not shown for the reduced model)} .}\par

\section*{\pr{O}rigin of the V-shaped C-N-H depletion pattern}

\begin{figure}[t]
 \centering
 \includegraphics[width=\linewidth]{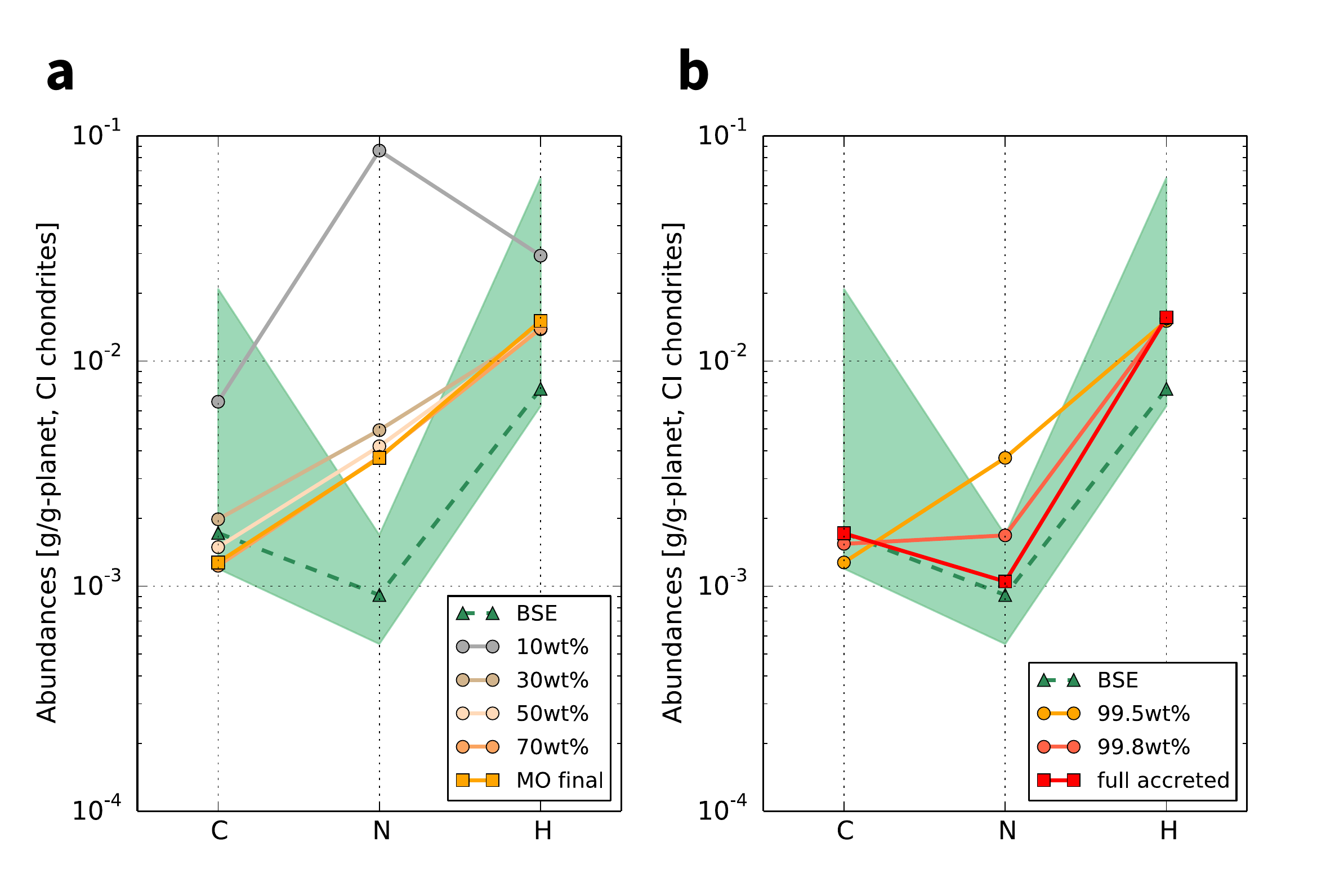}
 \caption{{\bf \pr{E}volution of major volatile abundances in \pr{the bulk silicate Earth (BSE)} scaled by those of CI chondrites in the nominal model.} The abundances are normalized by each planetary mass at each time for {\bf a,} the main accretion stage, from 10\% to 99.5\% of Earth's accretion, and {\bf b,} the late accretion stage defined as the last 0.5\% \pr{of} accretion after the magma ocean solidification. The time sequence is shown by lines from top to bottom with snapshots. The thick orange and red lines correspond to the end of main and late accretion stages, respectively. The range \prif{in the} current \pr{BSE} composition estimate\cite{hirschmann2016, hirschmann2018, marty2012} is shown for comparison (green area). The mean value of Hirschmann \cite{hirschmann2016} is shown as a reference for the relative depletion pattern (green line). See \tabref{tab:parameters} for the composition of BSE and chondrites.}  
 \label{fig:patternevolution} 
 \end{figure}

The V-shaped C-N-H depletion pattern compared \pr{with} chondrites in the current BSE \pr{(}\figref{fig:patternevolution}\pr{)} can be successfully reproduced from chondritic building blocks by considering both the element partitioning between reservoirs and the impact-induced atmospheric erosion simultaneously over \pr{all} accretion stages. 
The successful case (the nominal model) assumed three conditions: an oxidized magma ocean \HK{(\magenta{IW+1})}, \pr{a} change in impactor size distribution with time, and the storage of H and C in oceanic water and carbonate rocks after magma ocean solidification. The nominal model assumed CI chondrite-like building blocks, but \HK{enstatite} chondrite-like impactors' case\HK{, where a smaller amount of H is present in the building blocks,} also successfully reproduced \pr{the BSE} composition \red{under \HK{more} limited conditions} (see Supplementary Fig. S2, \pr{Fig.} S3\red{, Fig. S4, \pr{and Fig.} S5}). 
Element partitioning among the overlying atmosphere, magma ocean, and \pr{suspended metal droplets} during the main accretion \pr{phase} led to \pr{a} subtle C deficit, N excess, and an adequate amount of H at the end of the main accretion stage (\figref{fig:patternevolution}a). 
In the late accretion stage after solidification, the interplay of H and C storage in oceanic water and carbonate rocks and the preferential loss of N due to atmospheric erosion finally solved the remaining issue: N excess (\figref{fig:patternevolution}b).\par

\begin{figure}[ht]
 \centering
 \includegraphics[width=\linewidth]{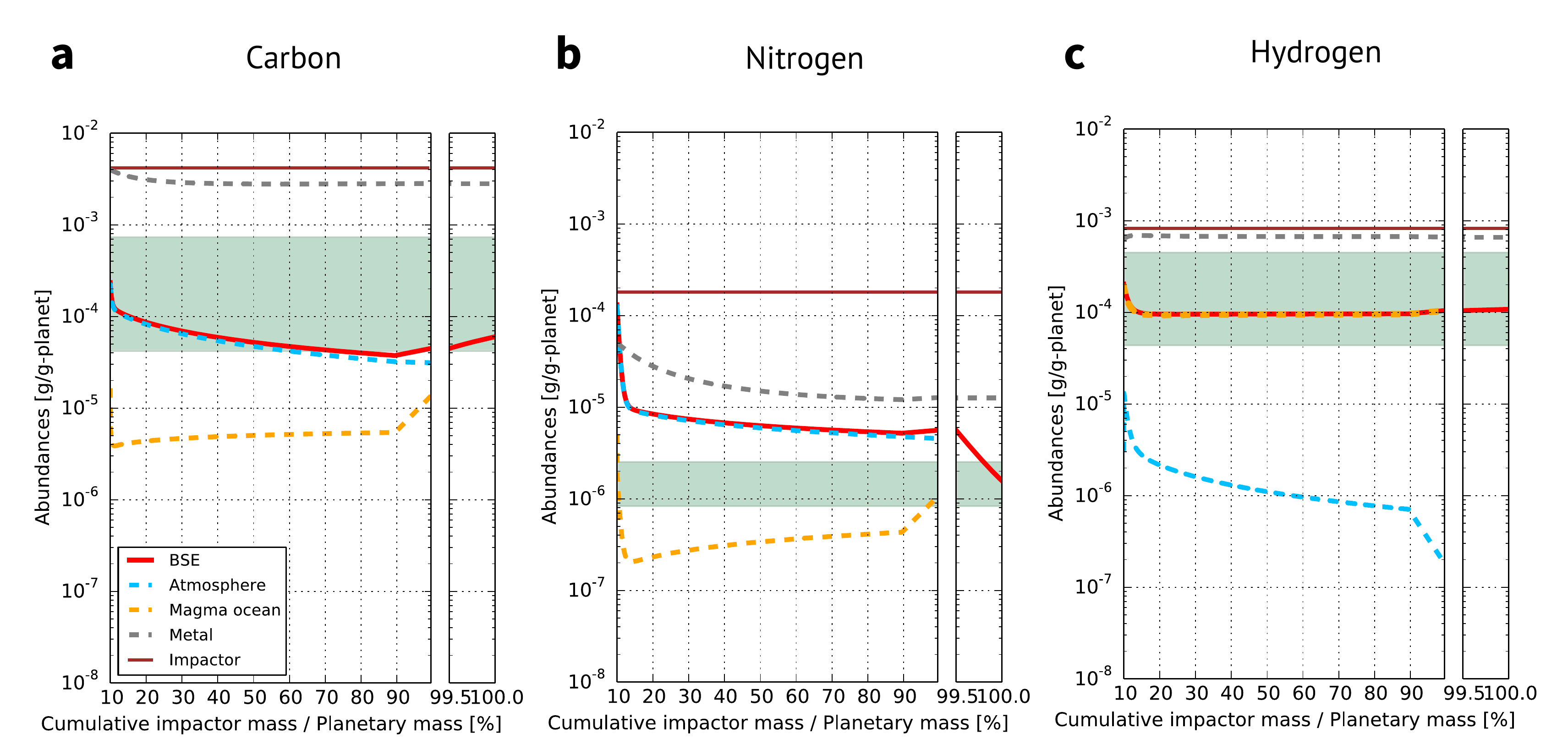}
 \caption{{\bf \pr{E}volution of the abundances of C, N, and H in the surface and interior reservoirs over the full accretion obtained from the nominal model}. Dashed lines correspond to the amounts in the atmosphere (light-blue), magma ocean (orange), and metallic core (grey) for the main accretion \pr{phase} and in the surface reservoirs (the atmosphere, oceans, and carbonate rocks\red{:} red solid line) for the late accretion \pr{phase}. \pr{S}olid lines mean the net cumulated into \pr{the bulk silicate Earth (BSE;} red) and delivered by impactors (brown). The green areas denote the amounts in the current BSE. Plotted abundances are scaled by the planetary mass at a given time. }
 \label{fig:CNHin3layer} 
 \end{figure}

Earth's C and H abundances were set chiefly during the main accretion stage (\figref{fig:CNHin3layer}). 
\pr{Alt}hough \pr{the} first kink of volatile abundances in each reservoir is set by \pr{the} initial condition\pr{s} (see Methods), the system soon evolves towards \pr{a} quasi-steady state between the gain and loss of volatile elements. 
The highly siderophile property of C\cite{dasgupta2013} and high solubility of H\cite{moore1998} in silicate melt under the oxidized condition in the nominal model caused those elements to be removed by core segregation.
\pr{As} the remaining part of C was partitioned into the atmosphere \pr{owing} to its low solubility\cite{pan1991}, the atmospheric erosion led to C being more depleted than H in BSE. The low solubility of N\cite{miyazaki2004} in magma \red{led} to almost all N in BSE \pr{being} partitioned into the atmosphere soon after the magma ocean solidification. 
For N, the impact-induced erosion governs the abundance evolution\cite{sakuraba2019}, while the transport to the core \pr{governs} for C and H. \red{N also \HK{has a} siderophile \HK{property}\pr{,} which \HK{has been proposed to cause} N depletion \pr{in the} \HK{BSE}\cite{dalou2017, grewal2019, grewal2021}. \HK{However,} we found that \HK{partitioning into the core is less important compared \pr{with}} the atmospheric escape\pr{,} \HK{even considering the uncertainty in the partitioning coefficient} (see Supplementary Information).} 
A Mars-sized, moon-forming giant impact\cite{canup2001} was assumed in the nominal model, which corresponds to the kink at 90\% Earth \pr{mass}, but it did not modify \pr{the BSE} volatile abundances significantly. 
We considered \HK{a} completely molten mantle\cite{canup2004} in the element partitioning after the giant impact, \red{while a smaller molten fraction of 30 wt.\% was assumed for the planetesimal accretion. \HK{Thus, the larger mass}} of the magma ocean \HK{after the giant impact} \red{allowed increases in the abundances of} all \HK{volatile elements} in the magma ocean.

The abundance of N was decreased by approximately one order of magnitude \pr{during} the late accretion \pr{phase owing} to impact-induced atmospheric escape. 
The formation of oceans and the initiation of the carbon\red{ate}-silicate cycle right after the solidification of the magma ocean trap\pr{ped} H and C into the surface reservoirs, and subsequently facilitate\pr{d} preferential N erosion from the atmosphere. 
Since N neither condense nor become incorporated into any solid or liquid reservoirs in our model, the final N abundance is determined by the balance between the supply by impactors and the loss by atmospheric erosion. \HK{From this result, we argue that} \red{the presence of oceans and \HK{carbonate formation} in the late accretion stage are requirements to explain the current high C/N and H/N ratio\pr{s of the BSE}. 

The exact timing of the ocean formation and the initiation of the \HK{carbonate}-silicate cycle are \HK{unknown}, but \HK{these assumptions are supported by theoretical, geological, and geochemical studies. \pr{As} the timescale for the magma ocean solidification is short} (0.2\HK{--}7 Myrs after the last giant impact\cite{salvador2017, nikolaou2019,hamano2013}), the oceans would have \HK{persisted} in the late accretion stage. Archean sediments imply that the \HK{oceans}\cite{appel1998} and plate tectonics\cite{komiya1999} already existed at least 3.8 Gyr ago. \HK{Geochemical studies} of Hadean zircons suggest the existence of \HK{oceans} and the active plate boundary interactions in \HK{the} Hadean\cite{wilde2001,hopkins2008}. } 
\pr{Alt}hough plate tectonics might not have started on the Hadean Earth\cite{korenaga2013}, the storage of C by carbonate precipitation is possible even without plate tectonics \red{as long as liquid water is present}\cite{foley2019}. 
\red{The continental crust\pr{,} which releases water-soluble cations such as Ca$^{2+}$ and Mg$^{2+}$ on modern Earth\pr{,} might not be present in the Hadean, but efficient carbonate formation is possible \prif{owing} to seafloor weathering \cite{krissansen2018}. However, the presence of liquid water which allows silicate weathering to occur is required to drive the carbonate\red{-silicate} cycle\cite{foley2015}.}
Furthermore, if marine pH was neutral to alkaline ($>\sim$7), most of the total atmosphere plus ocean C inventory ($>\sim$90\%) would \pr{have} dissolve\pr{d} into the oceans as bicarbonate and carbonate ions\cite{pierrehumbert2010}, as proposed for the preferential N loss by a giant impact\cite{tucker2014}. \red{We note\magenta{, however,} that a giant impact will vaporize the oceans, \HK{which} could end up losing some C.}

A\red{nother} key assumption of our model is the slow (negligible) N fixation \HK{compared to C} on early Earth during late accretion, which led to the preferential erosion of atmospheric N. 
A combined model of atmospheric and oceanic chemistry determined the lifetime of molecular N in anoxic atmospheres to be $>10^{9}$ years\cite{hu2019}. After Earth's accretion ceased, N cycling between the atmosphere and mantle \pr{over a} billion\pr{-}year timescale\cite{stueken2016} would lead to lower N partial pressures in the later period \pr{(e.g., those} recorded in the Archean\cite{catling2020}\pr{)}. 
In contrast, the timescale of carbonate precipitation even in the cation supply-limited regime ($\sim 10^6$ years\cite{foley2015}) is shorter than the duration of late accretion ($\sim 10^7$--$10^8$ years\cite{morbidelli2018}).

\section*{Size distribution of impactors and \pr{a}ccretion models} 

\begin{figure}[ht]
\centering
\includegraphics[width=13cm]{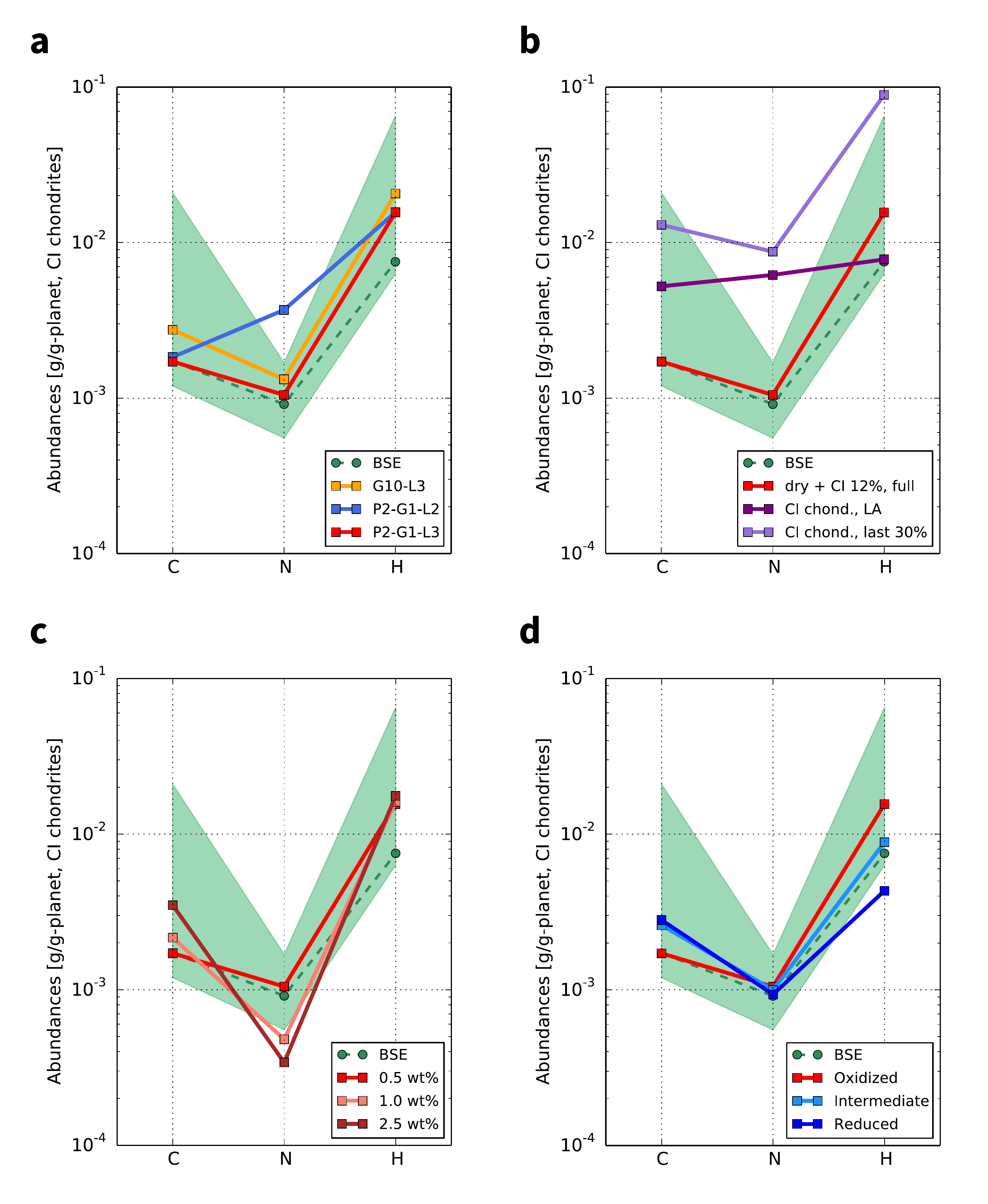}
 \caption{{\bf Dependence of final volatile composition of \pr{the bulk silicate Earth (BSE)} on the accreting conditions.} {\bf a)} \pr{E}ffects of the impactor's size distribution. P2-G1-L3 (the nominal model, red line): \pr{p}lanetesimal accretion ($q = 2$ and $q = 3$ for in the main and late accretion stages, respectively) and one giant impact. G10-L3 (orange line): \pr{t}en giant impacts and the late accretion of planetesimals . P2-G1-L2 (blue line): \pr{s}hallower planetesimal size distribution ($q = 2$ throughout the full accretion) and one giant impact. Here we assumed $\mathrm{d}N/\mathrm{d}D\propto D^{-q}$, where $N(D)$ is the number of objects of diameter smaller than $D$, and $q$ is the power law index. {\bf b)} \pr{D}ependence on volatile accretion scenarios. Late volatile accretion (dark-purple): volatiles are delivered only by \pr{late accretion} with CI chondrite-like bodies. Heterogeneous accretion model (purple): volatiles are supplied in the last 30 wt.\% accretion with CI chondrite-like bodies. {\bf c)} \pr{D}ependence on the late accretion mass. The mass of late accretion was varied from 0.5 wt.\% (brown) to 2.5 wt.\% (orange). {\bf d)} \pr{D}ependence on the redox state of the magma ocean. Oxidized (the nominal model, red line)\red{, intermediate (light blue line),} and reduced (blue line) conditions are compared. The solubilities and partitioning coefficients are summari\pr{s}ed in \tabref{tab:parameters}.}
 \label{fig:parametersv} 
 \end{figure}

Our results suggest the dominance of small \magenta{(km-sized)} impactors \pr{during} the late accretion. The nominal model assumed \pr{a} change in \pr{the} size distribution of impactors from shallower ($q = 2$ in $\mathrm{d}N/\mathrm{d}D\propto D^{-q}$, where $N(D)$ is the number of objects of diameter smaller than $D$, and $q$ is the power law index) for the main accretion \pr{phase} to steeper ($q = 3$, \pr{m}ain asteroid belt-like) for the late accretion \pr{phase} (\HK{see Methods}, \figref{fig:parametersv}a). 
The impact erosion by the late accretion which has a shallow size distribution is not \pr{sufficient} to reproduce \pr{the} BSE's N-depletion because km-sized small bodies are the most efficient in ejecting the atmosphere per unit mass of the impactor\cite{deNiem2012,schlichting2015}. 

The case where Earth formed by multiple giant impacts followed by \pr{late accretion} also reproduced the current \pr{BSE} volatile composition (\figref{fig:parametersv}a)\pr{; however}, \pr{atmospheric erosion} by small bodies during \pr{late accretion} is needed to obtain the V-shaped C-N-H pattern anyway. 
Since atmospheric loss per unit impactor mass is less efficient in giant impacts than in planetesimal accretion, \HK{larger} amount\red{s} of volatiles \pr{remained} in \pr{the} BSE. In addition, incomplete mixing between the impactor's core and magma ocean lower\pr{ed} the storage in the core. 

\red{The combination of a moon-forming giant impact with continuous planetesimal accretion in our nominal model is a plausible accretion scenario for Earth.} 
The dominance of planetesimal accretion and the change in the size distribution with time are naturally expected from the planet formation models. 
The giant impact stage \prif{was} set in the time when the total mass in planetesimals remain\prif{ed} comparable to the mass in protoplanets\cite{goldreich2004,kenyon2006,schlichting2015}. 
About half of giant impacts are not accretionary, but hit-and-run collisions which produce many fragments\cite{kokubo2010}. 
The change in size distribution from shallower to steeper is consistent with the inference that asteroids \pr{are} born big and evolved \pr{through} collisional cascade\pr{s}\cite{bottke2005, bottke2015}. 
\red{A \HK{size} distribution of late accretion impactors \HK{shallower than that of our nominal model ($q \sim 2$ vs. $q = 3$) has been proposed as the origin for the lunar} depletion \prred{of highly siderophile elements (HSE)} \HK{relative to Earth}\cite{bottke2010science}, but the \HK{long-lived} magma ocean \HK{on the Moon \pr{also} serves alternative explanation}\cite{morbidelli2018}.} 
Our scenario, in which Earth chiefly formed from a swarm of planetesimals, is in contrast to the pebble accretion model\cite{levison2015} and the idea \pr{that} attributes the Late Veneer to a single big impact\cite{brasser2016}, \pr{al}though impact fragments may act as small "planetesimals" to eject atmospheric N efficiently in those cases. 

\red{The timing of the volatile delivery onto the accreting terrestrial planets remains an open question\cite{albarede2009}.} We \red{assumed homogeneous accretion of CI chondritic building blocks combined with dry objects in the nominal model and} investigated the dependence on the variety of Earth's accretion models for both the volatile-rich late accretion\cite{chou1978, fischergodde2020} and the heterogeneous accretion model\cite{wanke1984,wade2005, schonbachler2010}  (\figref{fig:parametersv}b). \red{The latter two} models\pr{,} \red{which assumed later addition of 100\% CI chondrites accretion}\pr{,} result in much larger amount\red{s} of volatiles\red{, especially for N,} than the current \pr{BSE} inventory. \red{This means that \pr{the} volatile content fraction of the late accretion impactors appears more significant factor \pr{than} the total accumulated amount for explaining Earth's N depletion.} 
The uncertainty in the late accretion mass (0.5 wt.\% to 2.5 wt.\%\cite{marchi2018}) is considered in \figref{fig:parametersv}c. \prif{Greater} late accretion can erode more N and \pr{ac}cumulate more C, but V-shaped patterns were obtained over the range of mass \HK{estimates}. 

\section*{\prif{Magma ocean r}edox state}
\red{We explored how the redox state of the magma ocean affects the final volatile composition by considering oxidized (\magenta{IW+1}), intermediate (\magenta{IW-2}), and reduced (\magenta{IW-3.5}) conditions\cite{hirschmann2016} \HK{(see Methods)}.} The current C-N-H depletion pattern can be obtained under the oxidized \red{\HK{or} intermediate} magma ocean, \HK{while we ruled} out the reduced condition (\figref{fig:parametersv}d). \red{The redox state of the magma ocean\prif{,} \HK{which successfully reproduces \pr{the} BSE's abundance of major volatile elements}\prif{, has a} relatively oxidized condition ($\log_{10} f_{\rm O_2} \HK{\gtrsim} \magenta{\rm IW-2}$}). 
In the reduced model, the final amount of H corresponding to \magenta{1.15} ocean mass (0.1\magenta{5} ocean mass in the mantle) was obtained\pr{; this} is even smaller than the minimum estimate for present-day Earth’s mantle water content \pr{of} > 0.56 ocean masses \cite{korenaga2017, peslier2017, hirschmann2016, hirschmann2018, marty2012}. 
The behaviour of element partitioning during the main accretion stage is determined by \pr{solubilities in} the magma ocean and the partitioning coefficients between silicate melt and metal liquid, both of which depend on the pressure-temperature-$f_{\rm O_2}$ conditions of the magma ocean. 
According to what is currently understood, C becomes soluble and less siderophile, N becomes insoluble, and H becomes far more soluble in silicate liquids under \pr{oxidized conditions} compared \pr{with reduced conditions}\cite{hirschmann2016,tsuno2018,dalou2017,miyazaki2004,bureau1998}. 
\HK{We note that assuming enstatite chondrites as volatile sources \pr{fits only with the oxidized model} (Fig. \prred{S5}). } \par 

Since the flux from the atmosphere to the core is proportional to the product of solubility and partitioning coefficient \red{(see Method\pr{s})}, their influences on the resulting C cancelled each other out. The change in molecular masses (e.g., CO$_2$ = 44 amu in the oxidized model to CH$_4$ = 16 amu in the reduced model) also influences partial pressure and\magenta{, consequently,} effective solubility slightly, but the influence is not significant. 
For H, the storage in the magma ocean is important to reproduce the current \pr{BSE} abundance and so high solubility under oxidizing condition\pr{s} is required. We emphasize that even when we employed smaller values for the partitioning coefficient of H as argued by a few previous high-pressure metal-silicate partitioning experiments\cite{clesi2018,malavergne2019}, the final H amount does not increase under the \HK{reduced} magma ocean (see Supplementary Fig. \HK{S6}, case 'm') because the low solubility in silicate melt limits the influence of partitioning into metal on \pr{the} BSE's budget, supporting the idea of \red{a relatively} oxidized magma ocean during Earth's accretion.

The oxidizing magma ocean has been proposed by several studies and mechanisms to oxidize the magma ocean as Earth grew \pr{have been suggested\cite{siebert2013, badro2015,armstrong2019}}. 
Crystallization of perovskite at the depth of the lower mantle induces disproportion of \pr{ferrous} to ferric iron plus iron metal, the latter of which would then sink to the core, with the remnant in the mantle being oxidized (named self-oxidation\cite{wood2006}).
The pressure effect on a fixed Fe$^{3+}$/Fe$^{\rm total}$ ratio leads to a redox gradient where the surface becomes more oxidized than the depth\cite{hirschmann2012,armstrong2019}.
\pr{Both} are natural outcomes of the formation of an Earth-sized planet. 
\red{An increase in \HK{the mantle oxygen fugacity on growing Earth has been} proposed by many heterogeneous accretion models to explain Earth's bulk composition of refractory and moderately siderophile elements\cite{wade2005,wood2006,fischer2015,rubie2015}. 
As shown in \figref{fig:CNHin3layer}, the volatile distribution between the atmosphere, magma ocean, and metal soon converged towards \pr{quasi-steady state} balancing \pr{of} in/out fluxes of volatile elements within the first $\sim$20 wt.\% accretion. \HK{Given 20 wt.\% accretion being the mass required to reach the steady state for a give condition}, the requirement for the redox state of the magma ocean ($\log_{10} f_{\rm O_2} \gtrsim \magenta{\rm IW-2}$) \HK{should be \magenta{considered as that for}} the \HK{final} $\sim$20 wt.\% planetesimal accretion before the giant impact. This result does not contradict the initial reduced condition followed by a more oxidized state. 
}

\section*{\pr{Volatile p}redictions for Earth's core, bulk Venus\pr{,} and Mars}

Our scenario for the origin of Earth's volatile depletion pattern is testable with the further constraints of the composition of light elements in the core.
The final mass fractions in the metallic core in our nominal model, assuming an oxidized magma ocean \red{(\magenta{IW+1})}, were 0.9 wt.\%, 0.004 wt.\%, and 0.2 wt.\%for C, N, and, H, respectively. 
These predicted contents of light elements are within the range of each element's content allowance and \pr{account for approximately} 30\% of the Earth's core density \magenta{deficit}\cite{hirose2013}. 
Therefore, other light elements such as oxygen, silicon, and sulphur \pr{should also contribute} to the core density deficit. 
The \red{relatively} oxidized magma ocean required from our results may support the large contribution of oxygen\cite{badro2015}.
With upcoming data of solubilities and partitioning coefficients, this model \pr{will} provide a more accurate estimate of Earth's core composition (see also Supplementary Information).

Our model \pr{also} predicts the different depletion patterns of major volatile elements in bulk Venus and Mars.
Venus might never experience the condensation of liquid water and, consequently, \pr{carbonate precipitation} on the surface\cite{hamano2013}.
The lack of H and C storage leads to atmospheric loss of th\pr{e}se elements as well as N.
If \pr{atmospheric} CO$_2$ is the dominant C reservoir of bulk Venus, the total amount of C \prif{for} Venus is $\sim$0.4 times the value \prif{for} Earth\cite{dauphas2014}, supporting the model prediction. 
In contrast, the formation of H$_2$O and CO$_2$ ice on early Mars\cite{forget2013} allows the preferential loss of atmospheric N by impact erosion in the same manner as Earth\cite{sakuraba2019}. 
\pr{From the analysis of} SNC meteorites\cite{wadhwa2001, herd2002}\pr{, the} Martian mantle is thought to be very reduced with (\magenta{IW-1})\pr{,}  and previous models have predicted \red{a} more reduced magma ocean \prif{for} Mars than \prif{for} Earth\cite{hirschmann2008, zhang2017}. The reduced mantle would lead to a smaller storage of H in the magma ocean and \pr{a} subsequent deficit of bulk H.

\section*{Methods}

We calculated the abundances of C, N, and H in Earth growing from a Mars-sized embryo via accretion of planetesimals and protoplanets with a box model considering the partitioning between reservoirs and atmospheric erosion caused by impacts.
We defined two stages: main accretion and late accretion stages (\pr{e.g.,} the growth from 10\% to 99.5\% and the final 0.5\% of current Earth mass, respectively\pr{)} in our nominal model.
The former is also called the magma ocean stage.
For each finite mass step of planetesimal accretion, we solved the deterministic differential equations \pr{formally given} by,
\begin{equation}
    \frac{{\rm d} M_{\rm i}^{\rm BSE}}{{\rm d} M_{\rm p}} = x_{\rm i} - \sum_{\rm sinks} F^{\rm i}_{\rm k}, \label{eq:dMdM}
\end{equation}
where $M_{\rm i}^{\rm BSE}$, $M_{\rm p}$, $x_{\rm i}$, and $F^{\rm i}_{\rm k}$ are the total mass of element i in BSE, planetary mass, the abundance of volatiles in impactors, and outflux per unit mass accretion by the process k, respectively. As the volatile loss processes (sinks), we considered the atmospheric escape $F_{\rm esc}$ through the full accretion and the core segregation $F_{\rm core}$ only for the magma ocean stage. For each accretion step, the element partitioning between surface and interior reservoirs is calculated by the mass balance \pr{modelling}, 

\begin{eqnarray}
    M_{\rm i}^{\rm BSE} = \sum_{\rm j}M_{\rm i}^{\rm j},
\label{eq:Mitotal}
\end{eqnarray}
where the atmosphere, silicate melt, and suspended metal in the magma ocean in the main accretion stage, the atmosphere, ocean, and sedimentary carbonate rocks in the late accretion stage are considered as reservoirs j (see below for details). Each giant impact is treated separately from the statistically averaged planetesimal accretion. 
\red{\HK{We note that we confirmed numerical convergence by changing the step size of the cumulative accreted mass in our simulations.}} \par

{\bf Accretion model.} 
In our model, Earth grows by accreting planetesimals and protoplanets during the main accretion \pr{phase}, followed by \prif{late accretion} composed only of planetesimals. For planet growth, we consider\pr{ed} a change in \HK{the bulk} density caused by pressure by using Eq.\ \ref{eq:MRrelation}, which expresses the \red{mass-radius relationship for planets with an Earth-like composition. Seager et al.\cite{seager2007} \HK{showed a power-law relation} between the masses and radii of solid planets by model\pr{l}ing their interior structures. They provided fitted formulas for \HK{rocky planets with} 67.5 wt.\% MgSiO$_3$ + 32.5 wt.\% Fe as,}
\begin{eqnarray}
    \red{\log \tilde{R} = k_1+\frac{1}{3} \log \tilde{M} + k_2 \tilde{M}^{k_3},}
\label{eq:MRrelation}
\end{eqnarray}
\red{where $\tilde{R} = R/R_{\rm s}$ and $\tilde{M} = M/M_{\rm s}$ are the normalized radius and mass of terrestrial planets\pr{,} $k_{\rm i}$ is \pr{the} fitting constants (\HK{$k_1$} = 0.20945, \HK{$k_2 =$} 0.0804, \HK{$k_3 =$} 0.394), \pr{and} $r$ and $M_{\rm s}$ are the conversion factors \pr{obtained} as $R_{\rm s} = 3.19\ R_{\rm Earth}$ and $M_{\rm s} = 6.41\ M_{\rm Earth}$\pr{;} we used \HK{modified $R_{\rm s} = 3.29\ R_{\rm Earth}$ in our study to match Earth's mass and radius without changing the power-law index}.} 
The growth by planetesimal accretion \pr{was} investigated by a statistical method\cite{sakuraba2019} where the contribution of each impact \pr{was} averaged over their size and velocity distributions. 
The size distribution is given by a single power-law \prred{$\mathrm{d}N/\mathrm{d}D\propto D^{-q}$, where $N(D)$ is the number of objects of diameter smaller than $D$ and the index $q$ is a parameter. We assumed a shallow size distribution with $q = 2$ for the main accretion phase and a steep size distribution with $q = 3$, which corresponds to that of the present-day main belt asteroids\cite{bottke2005}, for the late accretion \pr{phase} in the nominal model, and investigated the dependence of the results by changing the power-law index (see main text).} 
The minimum and maximum sizes are assumed to be $10^{-1.5}$ km and $10^3$ km, respectively. 
We assumed a Rayleigh distribution for the random velocity of impactors as modelled by Sakuraba et al.\cite{sakuraba2019}, which corresponds to the Gaussian eccentricity distribution for impactors from the terrestrial planet feeding zone excited by a protoplanet\cite{idamakino1992}. 
Each giant impact is \HK{included} in a discrete way separately from continuous planetesimal accretion. \par

Volatiles have been delivered by the accreted bodies that have formed Earth. The building blocks of Earth are not fully understood and so we treat the abundance of volatiles in an impactor as \red{an unknown} parameter. 
We considered CI chondrites \HK{containing volatile elements which} accreted with dry planetesimals\cite{dasgupta2019}. 
CI chondrites contain 2--5 wt.\% C\cite{kerridge1985, pepin1991,gradywright2003}, 500--2000 ppm N\cite{kerridge1985, pepin1991,gradywright2003}, and 0.47--1.01 wt.\% ppm H\cite{kerridge1985,pepin1991,vacher2020}, and their atomic ratio of C/N is 17.0$\pm$3.0 \cite{bergin2015}. 
Enstatite chondrites contain 0.2--0.7 wt.\% C\cite{pepin1991, gradywright2003}, 100--500 ppm N\cite{gradywright2003}, and 90--600 ppm H\cite{robert2003,piani2020}, and their C/N is 13.7$\pm$ 12.1\cite{bergin2015}. In our model, we assume\pr{d} the reference abundances as listed in \tabref{tab:parameters}. 
The fraction of CI chondrites was used as a parameter and set to be 12 wt.\% in the nominal model (Supplementary Fig. \prred{S2}). 
The results for \HK{cases where} enstatite chondritic impactor\HK{s are considered as volatile sources} are also shown in the Supplementary Information (Supplementary Fig. \prred{S3 and \pr{Fig.} S4}).

Giant impactors would have experienced core-mantle differentiation and volatile loss by atmospheric erosion. 
We calculated the abundances of C, N, and H in protoplanets by running our model for the growth by planetesimal accretion in advance from 0.05 to 0.1 Earth masses and then adapted the result to the compositions of giant impactors.

\HK{We assume\prred{d} that 32.5 wt.\% of the impactor mass is added to Earth as metallic iron regardless of the impactor \pr{type} to reproduce the mass fraction of Earth's core. The metal mass fraction is not necessarily equal to that of impactors because the former would be controlled by redox reactions (namely, the oxygen fugacity of the magma ocean), which are not explicitly modelled in our study}. 

{\bf Atmospheric erosion and loss of impactors.} 

Atmospheric erosion and loss of the impactors themselves through planetesimal accretion are computed as compiled by Sakuraba et al.\cite{sakuraba2019}. The net impact-induced escaping flux per unit impactor mass is given by
\begin{eqnarray}
    F_{\rm esc}=\zeta x_{\rm i}+\eta\label{eq:model_AE},
\end{eqnarray}
where $x_{\rm i}$ is the abundance of volatiles in the impactor, $\eta$ is the erosion efficiency of the atmosphere, and $\zeta$ is the escaping efficiency of the impactor vapor, and calculated for each element i. \red{The \HK{erosion and} escaping \HK{efficiencies for a single impact} can be estimated from input impactor's size and \pr{impact velocity}. We calculated the efficiencies \HK{statistically-averaged} over both impactor's size and velocity \HK{distributions} assumed in the above accretion model.} 
\red{We adopted the scaling laws obtained by numerical simulations for the atmospheric escape and for loss of an impactor's vapour.} Realistic numerical simulations of atmospheric erosion were given in 3D geometry by Shuvalov\cite{shuvalov2009}, and in 2D cylindrical geometry by Svetsov\cite{svetsov2007}. 
In Sakuraba et al.\cite{sakuraba2019}, \pr{the} Svetsov model\cite{svetsov2000, svetsov2007} and Shuvalov model\cite{shuvalov2009} are adopted for the atmospheric erosion efficiency $\eta$ (Equation 5 in Sakuraba et al.\cite{sakuraba2019}) and the impactor's escaping efficiency $\zeta$ (Equation 8 in Sakuraba et al.\cite{sakuraba2019}), respect\pr{i}vely. \red{The effect of oblique impacts that \pr{enhance} the impact erosion is considered by including the angle-averaged factor estimated by Shuvalov model\cite{shuvalov2009}.}
The volatiles in the fraction of impactors \pr{that} avoided the loss are released to the atmosphere and then partitioned into other reservoirs. 
Since the atmospheric loss mass is proportional to the abundance of each species in the atmosphere, element partitioning between the atmosphere and the other solid or liquid reservoirs is important for atmospheric erosion\cite{sakuraba2019}. 
We used the surface temperatures, 1,500 K for the main accretion stage and 288 K for the late accretion stage, given below to compute the atmospheric density and scale height, but we confirmed that the results are insensitive to the atmospheric temperature (Supplementary Fig. \prred{S1a}). \par

For each giant impact, we calculated the atmospheric loss from the mixture of the proto-atmosphere and the impactor's atmosphere caused by the global ground motion by using the model of Schlichting et al.\cite{schlichting2015}. 
We assumed Mars-sized (0.1 Earth masses) impactors whose impact velocity is 1.1 times the mutual escape velocity as commonly considered for the Moon-forming impact (see, for example, Hosono et al.\cite{hosono2019}). \red{We note that the estimates for the giant impact velocity has uncertainty from 1.0 to 1.2 times the escape velocity\cite{kaib2015}.} 
We \HK{also} note that recent 3D smoothed particle hydrodynamics simulations\cite{kegerreis2019, kegerreis2020} suggest a lower erosion efficiency than that of Schlichting model by a fac\pr{t}or of 3. 
\red{However, we found that the dependence on the atmospheric erosion efficiency by a giant impact \pr{was negligible} even if the atmospheric erosion was more or less efficient by a factor of \HK{\pr{three}} (Supplementary Fig. \prred{S1b}) \HK{\prif{owing to} the range \prif{in}} uncertainty \HK{aris\prif{ing} from} the velocity of a giant impact \HK{\prif{and/}or} the model dependency.}

{\bf Core segregation.} 
We considered the core segregation under the solidified mantle in the magma ocean stage by excluding the segregated metal from the equilibrium partitioning calculation. \pr{C}ore-forming liquid metal droplets carry volatile elements from the magma ocean to the metal pond, and consequently, to the core. We assumed the suspended metal fraction in the magma ocean to be constant\pr{;} then\pr{,} the net segregation flux per unit impactor mass is given by 
\begin{eqnarray}
    F_{\rm core}= \Cimet \left(x_{\rm met} - X_{\rm MO}\Xmomet\right)\label{eq:model_CS},
\end{eqnarray}
where $\Cimet$, $x_{\rm met}$, $X_{\rm MO}$, and $\Xmomet$ are the volatile concentration in the metal, the metal fraction of the \HK{newly accreted mass (assumed to be 32.5 wt.\%)}, the melting fraction of the planet, and the metal mass fraction in the magma ocean, respectively. \red{Our model is based upon previous studies\cite{bergin2015,hirschmann2016} \pr{that} assumed that core-mantle separation is a single stage event as a simplification, but we improved the model to track time evolution through accretion, where core formation is a contentious process.}\par

\pr{During} growth by planetesimal accretion, the mass fraction of \pr{the} molten magma ocean $X_{\rm MO}$ \pr{was} fixed to 30 wt.\% of the planetary mass\prnc{.} This \pr{was} derived from the estimated depth of \pr{the} magma ocean (30\pr{\%}-40\% of the mantle) constrained from the abundance of refractory siderophile elements (Ni and Co) in the present-day mantle\cite{wood2006}. A deeper magma ocean is also suggested\cite{fischer2015}, but we confirmed that the results do not change significantly even if we consider a deeper magma ocean \pr{of} up to 60 wt.\% (see Supplementary Fig. \prred{S1c}).\par

\red{As the Earth \HK{grows}, the metal droplets \HK{descend} through the deep magma ocean, continuously equilibrating with the silicate liquid\cite{rubie2003}. Metal droplets \pr{that} have reached the base of \red{\pr{the}} magma ocean forms metal pond\pr{s} and subsequently descend \HK{as} large diapirs to the growing core without further equilibration with the surrounding silicate\cite{wood2006}.} The metal in the magma ocean is assumed to be in equilibrium with the entire magma ocean and the mass fraction of the suspended metal $\Xmomet$ is fixed to $10^{-6}$. This reference value is estimated from the typical timescales of metal droplets settling ($\train\sim 10^1$ yrs\cite{rubie2015treatise}) and accretion of \pr{the} Earth ($\tacc\sim 10^7$ yrs\cite{kleine2002}) by, 
 \begin{eqnarray}
    \Xmomet \sim (1-\zeta)\Xmet M_\oplus\cdot\frac{\train}{\tacc} \sim 10^{-6}.
 \label{eq:Mmomet}
 \end{eqnarray}
\red{The settling timescale \HK{of the metal droplets was} estimated from the magma ocean depth $\sim$10$^2$--10$^3$ km divided by \pr{the} settling velocity $\sim$0.5 cm/s\cite{rubie2003}. In fact it evolves with time, but we confirmed} that the results do not depend on the suspended metal fraction 
unless it exceeds 10$^{-2}$ because metal droplets carry \red{almost} \pr{the} same mass of volatiles \red{in total} anyway (Supplementary Fig. \HK{\prred{S1d}}). \red{The sum of remaining fraction of metal is assumed to have segregated from the magma ocean into the metal pond and, subsequently, into the core \pr{and} are treated as one isolated reservoir}. \par 

In contrast to planetesimal accretion, giant impacts lead to a deeper magma ocean \HK{and} incomplete mixing between the magma ocean and the core of the impactor.
Since the Mars-sized moon-forming giant impact is thought to have delivered enough energy to completely melt the whole mantle\cite{pritchard2000}, we assumed a fully molten mantle after the giant impacts (see, for example, \HK{Canup}\cite{canup2004}). Moreover, we confirmed that assuming partial melting for a giant impact does not change our results significantly (Supplementary Fig. \HK{\prred{S1c}}). 
For \prnc{metal--silicate mixing} in re-equilibration between the impactor's core and proto-magma ocean, we considered \prnc{turbulent entrainment} by applying the formula of the metal fraction in the metal-silicate mixture to the completely molten mantle (Equation 6 in Degen et al.\cite{deguen2011}). 
 
The solidified mantle is assumed to be volatile free and is not considered as a reservoir because of its low capacity compared \pr{with} the magma ocean (see, for example, \HK{Elkins-Tanton}\cite{elkins2008}). \red{\prred{In terms of} H and C, it is suggested that considerable amount\HK{s} could be retained in the residual solid mantle as melt inclusions upon magma ocean solidification\cite{hier2017}, but \pr{this} would not affect the final \pr{BSE} inventory estimates.} \pr{While} we \pr{did} not consider the volatile trapping by \pr{the} solid silicate reservoir during the crystallization of the magma ocean, we \pr{did} \red{consider \HK{efficient} trapping \pr{of} outgassed H and C into the ocean\HK{s} and \HK{the crustal carbonate reservoirs as well as the mantle, respectively,} which \HK{are} included as BSE abundances. Whether H and C are trapped in the mantle or in the surface reservoirs does not affect the evolution of \pr{the BSE} volatile contents.} In the case of N, \red{since the \HK{N} partitioning \HK{coefficient} between mantle minerals and silicate melt is smaller than unity by orders of magnitude even under high temperature\cite{li2013}, almost \pr{all} \HK{N} would have been enriched in the melt during the crystallization and subsequently outgassed \pr{to form} the early atmosphere. Hier-Majumder and Hirschmann\cite{hier2017} \HK{showed} that owing to extremely low solubility of N$_2$, N retention into the residual mantle \HK{is inefficient}. 
For these reasons, the incorporation of volatiles into the solidified mantle does not change our conclusions.}

{\bf Equilibrium partitioning in magma ocean.} 
We calculated the partitioning of elements between the magma ocean, core-forming alloy, and overlying atmosphere (\figref{fig:modelimage}a), assuming \prnc{equilibrium partitioning}. \red{For the element partitioning between these three reservoirs, the mass balance equation (Eq. \ref{eq:Mitotal}) can be written as, 
\begin{eqnarray}
    M_{\rm i}^{\rm BSE} =\Miatm + \Misil + \Mimet=\frac{4\pi R^2\miatm}{g\bar{m}r_{\rm i}}P_{\rm i}+\Cisil\Msil+\Cimet\Mmet,
\label{eq:Mitotal_MO}
\end{eqnarray}
where $P_i$ is the partial pressures, $g$ is the gravitational acceleration \HK{given} by $g = \frac{GM}{R^2}$\prred{,} \HK{where} $G$ is the \HK{gravitational} constant, $\bar{m}$ is the mean molecular mass of the atmosphere, $\miatm$ is the molecular mass of the volatile species, $r_{\rm i} = \miatm/m_{\rm i}$ is the mass ratio of the volatile species to the element of interest (e.g., $r_{\rm C} = {m_{\rm CO_2}}/{m_{\rm C}} = 44/12$ for C, $r_{\rm N} = {m_{\rm N_2}}/2{m_{\rm N}} = 1$ for N, $r_{\rm H} = {m_{\rm H_2O}}/2{m_{\rm H}} = 18/2$ for H in the oxidized condition), $C_{\rm i}^{\rm j} = M_{\rm i}^{\rm j}/M_{\rm j}$ is the \magenta{mass} concentration in the silicate and metal, and $\Msil$ and $\Mmet$ are the mass\HK{es} of the magma ocean and the metal liquid in the magma ocean.}
The partition coefficient \HK{which} characterize\HK{s} \HK{elemental partitioning} between silicate and metal \prif{can be written as}, 
\begin{eqnarray}
\Dims = \frac{\Cimet}{\Cisil}.\label{eq:Dims}
\end{eqnarray} 
Equilibrium between the atmosphere and the underlying silicate melt is given by a solubility law, that in many cases can be approximated by a Henrian constant, 
\begin{eqnarray}
S_i\ (x = 1, 2) = \red{\frac{\Cisil}{P_i^{\frac1x}}}\label{eq:SiCisil},
\end{eqnarray}
\HK{where $x$ is the ratio of the number of atoms between gas and solute phases for the element of interest} (e.g., $x = 1$ for C and N, $x = 2$ for H, \HK{see below}). 

We \HK{defined} \red{three} models for the redox state of the magma ocean\HK{:} \red{the oxidized (\magenta{$\log_{10} f_{\rm O_2} \sim \magenta{\rm IW+1}$}), intermediate (\magenta{IW-2}), and  reduced (\magenta{IW-3.5}) conditions\HK{\cite{hirschmann2016}}\magenta{, where $f_{\rm O_2}$ is the oxygen fugacity\magenta{, ${\rm IW}$ is defined hereafter as $\log_{10} f_{\rm O_2}^{\rm IW}$, and $f_{\rm O_2}^{\rm IW}$ is} $f_{\rm O_2}$ \magenta{at} the iron-w\"{u}stite buffer}.} \red{We call the redox state of $\log_{10} f_{\rm O_2} \sim \log_{10} f_{\rm O_2}^{\rm IW} -2$ \prif{an} 'intermediate' condition because it \HK{corresponds} to that estimated for a single stage core formation scenario\cite{palme2014}.} As proposed by several studies\cite{armstrong2019, badro2015}, we assumed \red{an} oxidized magma ocean in the nominal model, \HK{and investigated the dependence of the results by changing the redox state}. \red{\HK{The conditions for} the metal-silicate \HK{equilibrium} at the bottom of \HK{Earth's} magma ocean \prred{were} estimated to be \HK{at} $\sim$40 GPa from Ni and Co partitioning, and \HK{at} $\sim$3750 K from V and Cr partitioning\cite{wade2005}.} \red{We assumed the partitioning coefficients thought to be applicable to \HK{these high $P$-$T$ conditions mentioned above for each redox state model} (ref.\cite{chi2014,armstrong2015,okuchi1997,roskosz2013}) as tabulated by Hirschmann \cite{hirschmann2016} (see \tabref{tab:parameters}).} 
Since the dependence of \HK{the partitioning coefficients} on $P$-$T$-$f_{\rm O_2}$ \HK{conditions} \HK{has not been fully understood}, we \HK{additionally} investigated the sensitivity of our model by varying partitioning coefficients for a wide range \HK{suggested from the literature:} $\Dcms$ = \red{0.5--5670\cite{dasgupta2013, chi2014, armstrong2015, li2016carbon, dalou2017, tsuno2018, malavergne2019, fischer2020}}, 
$\Dnms$ = \red{0.003--150\cite{roskosz2013, li2016nitrogen, dalou2017,speelmanns2019, grewal2019fate, grewal2019, grewal2021}}
, and $\Dhms$ = \red{0.2--100}\cite{okuchi1997,shibazaki2009, clesi2018, malavergne2019, li2020, tagawa2021}
(\HK{Supplementary Text, Supplementary Tab. S1 and} Supplementary Fig. \prred{S6}).\par

We considered different atmospheric components and used the constant solubilities and partitioning coefficients \red{for each \HK{redox state} model:} CO$_2$, N$_2$, and H$_2$O for the nominal oxidized model\pr{;} \red{CO, N$_2$, and H$_2$ for the intermediate model\pr{;}} and CH$_4$, NH$_3$, and H$_2$ for the reduced model. \red{All these species of molecules in the atmosphere are assumed by following Hirschmann\cite{hirschmann2016}}.
As summ\red{a}ri\pr{s}ed in \tabref{tab:parameters}, we fixed solubilities for C and N and used \pr{the} Moore model\cite{moore1998} for H by following Hirschman\cite{hirschmann2016}. \HK{Considering the range of solubilities reported by} experimental \HK{studies}, we \HK{also tested} the model sensitivity for $S_{\rm C}$ = (\red{0.002--8})\ $\left(\frac{P_{\rm CO_2}}{\rm MPa}\right)$ ppm\cite{pan1991,bureau1998,wetzel2013,stanley2014,armstrong2015}, $S_{\rm N}$ = (\red{0.2--46})\ $\left(\frac{P_{\rm N_2}}{\rm MPa}\right)$ ppm\cite{miyazaki2004, libourel2003, boulliung2020}. For $S_{\rm H}$, we tested for more and less soluble cases than \pr{the} Moore model by a factor of \pr{two} (\HK{Supplementary Text,} Supplementary Tab. S1, and Fig. S\HK{\prred{6}}). \red{By substituting Eq. (\ref{eq:Dims}) and (\ref{eq:SiCisil}) to Eq. (\ref{eq:model_CS}), we can get the relationship of $F_{\rm core}\propto \Cimet \propto S_{\rm i}\Dims$ which means \HK{that} the volatile flux of core segregation is proportional to the product of solubility and \prif{the} partitioning coefficient of each element i\cite{hirschmann2016}.}\par

{\bf \pr{Late accretion stage surface} environment.} 
As we defined the late accretion as the last 0.5 wt.\% accretion, we examined whether the surface \pr{could} melt again after the magma ocean solidification by roughly estimating the energy balance between the gravitational accretion and the planetary radiation from the oceanic surface during \prif{late accretion}. 
The input accretional energy is estimated by $E_{\rm in} = (GM_{\rm Earth}M_{\rm LA}/R_{\rm Earth})\tau_{\rm LA}^{-1}$, where $M_{\rm LA}$ is 0.5 wt.\% of \pr{the} Earth mass and $\tau_{\rm LA}\sim10^8$ yrs\cite{bottke2010science} is the timescale of the late accretion. The maximum outgoing planetary radiation flux is estimated by assuming the radiation limit from the saturated atmosphere ($\sim$300 W/m$^2$\cite{abe1988,kasting1988,nakajima1992}). 
As a result, the latter exceeded the former by $\sim$2 orders of magnitude \pr{and} confirmed that the planetary surface would not melt again during the late accretion. \par 

We assumed the formation of the oceans and onset of the \HK{carbonate}-silicate cycle (\figref{fig:modelimage}b) as modelled by Sakuraba et al.\cite{sakuraba2019}. \red{For the surface conditions, \prif{it is suggested that} \HK{the oceans have} form\prif{ed} \prred{within} the several Myrs \HK{after the Moon-forming giant impact} by \HK{calculating} the timescale of the magma ocean solidification \cite{salvador2017, nikolaou2019, hamano2013}. Furthermore, \HK{voluminous} oceans underlain by a dry mantle\pr{,} like that right after the magma ocean solidification\prif{, create an} ideal situation to drive plate tectonics\cite{korenaga2013,kurokawa2018subduction}.} 
\red{\HK{The carbonate}-silicate cycle is believed to have stabilized the Earth’s climate \pr{over} a long geological timescale against the increase \pr{in} solar luminosity over the past 4 billion years\cite{walker1981}. As long as its negative feedbacks of temperature-dependent continental and seafloor weathering has been driven throughout Earth history as assumed here, the early climate was likely temperate\pr{, with} temperature\pr{s} on Earth is estimated to be 273 K to 370 K\cite{kasting1993,krissansen2018}}. \red{Since we confirmed \pr{that} dependence on the temperature within this range is negligible \magenta{(\HKcomment{the figure} not shown)}, w}e assume\pr{d} \red{the present-day global mean temperature} \pr{of} $T$ = 288 K \red{as the reference value.} 
We calculated the partitioning of elements between the atmosphere, oceans, and crust plus mantle.  
The vapour-liquid equilibrium sets an upper limit to the partial pressure of H$_2$O in the atmosphere, $P^{\rm crit}_{\rm H_2O} = 1.7\times10^{-2}\ {\rm bar}$\cite{murphykoop2005}, calculated for the assumed surface temperature \pr{of} 288 K. 
We impose the negative feedback of the carbon\red{ate}-silicate cycle by simply setting an upper limit to the partial pressure of $P^{\rm crit}_{\rm CO_2} = 10$ bar, as expected for the steady state \cite{kasting1993}. 
Neglecting the time lag to reach the steady state is justified by considering the short timescale of carbonate precipitation compared \pr{with} the duration of late accretion (see main text).
Atmospheric H and C in excess from the upper limits are partitioned into oceans and crust plus mantle reservoirs, respectively.

{\bf Initial condition.} 
We prepared the initial condition for the elemental abundances by assuming a chondritic bulk composition the same as planetesimal impactors and equilibrium partitioning between the atmosphere, the fully molten magma ocean, and the core.
\pr{While} this is a crude assumption \pr{that neglects} the complexity of how planetary embryos formed, we have confirmed that the results are insensitive to the initial condition because the system soon evolves towards the quasi-steady state between the gain and loss of volatile elements. 

\begin{table}[ht]
\centering
\small
\caption{\pr{L}ist of key parameters for the model \red{calculations}. Observational data and model reference values for the abundances of C, N, and H in \pr{the bulk silicate Earth (BSE)} and chondrites (assumed as building blocks, CI chondrites and \HK{enstatite} chondrites). We set CI chondritic model, solubilities $S_{\rm i}$ (the units are $({P_{\rm i}}/{\rm 1 MPa})^{1/2} \ {\rm ppm}$ for reduced $S_{\rm H}$ and $\left({P_{\rm i}}/{\rm 1 MPa}\right) \ {\rm ppm}$ for the others, where $P_{\rm i}$ is the partial pressure of each molecule ${\rm i}$), and the partitioning coefficients $D_{\rm i}$ under each redox state by following Hirschmann (2016)\cite{hirschmann2016}. a: Hirschmann (2016)\cite{hirschmann2016}, b: Marty (2012)\cite{marty2012}, c: Hirschmann (2018)\cite{hirschmann2018}, d: Bergin et al. (2015)\cite{bergin2015}, e: Hirschmann \& Dasgupta (2009), f: Grady \& Wright, (2003)\cite{gradywright2003}, g: Vacher et al. (2020)\cite{vacher2020}, h: Kerridge (1985) \cite{kerridge1985}, i: Piani et al. (2020)\cite{piani2020}\prred{, j: Stolper \& Holloway (1988)\cite{stolper1988}, k:Pan et al. (1991)\cite{pan1991},  l:Libourel et al. (2003)\cite{libourel2003} m:Moore et al. (1998) \cite{moore1998}, n:Chi et al. (2014)\cite{chi2014}, o:Armstrong et al. (2015)\cite{armstrong2015}, p:Roskosz et al. (2013)\cite{roskosz2013}, q:Okuchi, (1997)\cite{okuchi1997}, r: Wetzel et al. (2013)\cite{wetzel2013} s:Hirschmann et al. (2012)\cite{hirschmann2012}.}}
  \begin{tabular}{rccccc}\hline 
    & C  & N & H & C/N & C/H \\ \hline
   BSE [$\mu$g/g-Earth] & 42--730$^{\rm a,b,c}$ & 0.83--2.5$^{\rm a,b,c}$ & 44--450$^{\rm a,b,c}$ & 40 $\pm$ 8$^{\rm a,d}$ & 1.3 $\pm$ 0.3 $^{\rm a,e}$\\
   CI chondrites [ppm] & 20,000--50,000$^{\rm f}$ & 500--2,000 $^{\rm f}$ & 4,700--10,500$^{\rm g}$ & 14.5 $\pm$ 2.5$^{\rm d}$ & 2--8$^{\rm h}$ \\ 
   CI model [ppm] & 35,000 & 1,500 & 6,900 & 23 & 5\\
   Enstatite chondrites [ppm] & 2,000--7,000$^{\rm f}$ & 100--500 $^{\rm f}$ & 90--600 $^{\rm i}$ & 13.7$\pm$ 12.1$^{\rm d}$ & \\
   EC model [ppm] & 4,000 & 250 & 400 & 16 & 10\\ \hline
   Oxidised model& & & & & \\ %
   $S_{\rm i}$ & 1.6\red{$^{\rm j,k}$} & 1.0\red{$^{\rm l}$} & M98 model\red{$^{\rm m}$} & & \\ 
   $D_{\rm i}^{\rm met/sil}$ & 500\red{$^{\rm n,o}$}  & 20\red{$^{\rm p}$} & 6.5\red{$^{\rm q}$} & & \\
   \red{Intermediate model}& & & & & \\ %
   \red{$S_{\rm i}$} & \red{0.55$^{\rm o}$} & \red{5.0$^{\rm p}$} & \red{M98 model$^{\rm l}$} & & \\ 
   \red{$D_{\rm i}^{\rm met/sil}$} & \red{1000$^{\rm n,o}$}  & \red{20$^{\rm p}$} & \red{6.5$^{\rm q}$} & & \\
   Reduced model & & & & & \\
   $S_{\rm i}$ & 0.22\red{$^{\rm r}$} & 50\red{$^{\rm l}$} & 5.0\red{$^{\rm s}$} & & \\
   $D_{\rm i}^{\rm met/sil}$ & 3,000\red{$^{\rm n,o}$}  & 20\red{$^{\rm p}$} & 6.5\red{$^{\rm q}$} & & \\
   \hline
  \end{tabular}
    \label{tab:parameters}
\end{table}

\bibliography{mybib}

\begin{thebibliography}{100}
\urlstyle{rm}
\expandafter\ifx\csname url\endcsname\relax
  \def\url#1{\texttt{#1}}\fi
\expandafter\ifx\csname urlprefix\endcsname\relax\def\urlprefix{URL }\fi
\expandafter\ifx\csname doiprefix\endcsname\relax\def\doiprefix{DOI: }\fi
\providecommand{\bibinfo}[2]{#2}
\providecommand{\eprint}[2][]{\url{#2}}

\bibitem{abbot2012}
\bibinfo{author}{Abbot, D.~S.}, \bibinfo{author}{Cowan, N.~B.} \&
  \bibinfo{author}{Ciesla, F.~J.}
\newblock \bibinfo{journal}{\bibinfo{title}{Indication of insensitivity of
  planetary weathering behavior and habitable zone to surface land fraction}}.
\newblock {\emph{\JournalTitle{The Astrophysical Journal}}}
  \textbf{\bibinfo{volume}{756}}, \bibinfo{pages}{178} (\bibinfo{year}{2012}).

\bibitem{wordsworth2013}
\bibinfo{author}{Wordsworth, R.} \& \bibinfo{author}{Pierrehumbert, R.}
\newblock \bibinfo{journal}{\bibinfo{title}{Hydrogen-nitrogen greenhouse
  warming in {E}arth's early atmosphere}}.
\newblock {\emph{\JournalTitle{science}}} \textbf{\bibinfo{volume}{339}},
  \bibinfo{pages}{64--67} (\bibinfo{year}{2013}).

\bibitem{foley2015}
\bibinfo{author}{Foley, B.~J.}
\newblock \bibinfo{journal}{\bibinfo{title}{The role of plate tectonic--climate
  coupling and exposed land area in the development of habitable climates on
  rocky planets}}.
\newblock {\emph{\JournalTitle{The Astrophysical Journal}}}
  \textbf{\bibinfo{volume}{812}}, \bibinfo{pages}{36} (\bibinfo{year}{2015}).

\bibitem{marty2012}
\bibinfo{author}{Marty, B.}
\newblock \bibinfo{journal}{\bibinfo{title}{The origins and concentrations of
  water, carbon, nitrogen and noble gases on {E}arth}}.
\newblock {\emph{\JournalTitle{Earth and Planetary Science Letters}}}
  \textbf{\bibinfo{volume}{313}}, \bibinfo{pages}{56--66}
  (\bibinfo{year}{2012}).

\bibitem{hirschmann2016}
\bibinfo{author}{Hirschmann, M.~M.}
\newblock \bibinfo{journal}{\bibinfo{title}{Constraints on the early delivery
  and fractionation of {E}arth's major volatiles from {C/H, C/N}, and {C/S}
  ratios}}.
\newblock {\emph{\JournalTitle{American Mineralogist}}}
  \textbf{\bibinfo{volume}{101}}, \bibinfo{pages}{540--553}
  (\bibinfo{year}{2016}).

\bibitem{halliday2013}
\bibinfo{author}{Halliday, A.~N.}
\newblock \bibinfo{journal}{\bibinfo{title}{The origins of volatiles in the
  terrestrial planets}}.
\newblock {\emph{\JournalTitle{Geochimica et Cosmochimica Acta}}}
  \textbf{\bibinfo{volume}{105}}, \bibinfo{pages}{146--171}
  (\bibinfo{year}{2013}).

\bibitem{walsh2011}
\bibinfo{author}{Walsh, K.~J.}, \bibinfo{author}{Morbidelli, A.},
  \bibinfo{author}{Raymond, S.~N.}, \bibinfo{author}{O'brien, D.~P.} \&
  \bibinfo{author}{Mandell, A.~M.}
\newblock \bibinfo{journal}{\bibinfo{title}{A low mass for {M}ars from
  {J}upiter{'}s early gas-driven migration}}.
\newblock {\emph{\JournalTitle{Nature}}} \textbf{\bibinfo{volume}{475}},
  \bibinfo{pages}{206--209} (\bibinfo{year}{2011}).

\bibitem{dauphas2017}
\bibinfo{author}{Dauphas, N.}
\newblock \bibinfo{journal}{\bibinfo{title}{The isotopic nature of the
  {E}arth's accreting material through time}}.
\newblock {\emph{\JournalTitle{Nature}}} \textbf{\bibinfo{volume}{541}},
  \bibinfo{pages}{521--524} (\bibinfo{year}{2017}).

\bibitem{fischergodde2020}
\bibinfo{author}{Fischer-G{\"o}dde, M.} \emph{et~al.}
\newblock \bibinfo{journal}{\bibinfo{title}{Ruthenium isotope vestige of
  {E}arth's pre-late-veneer mantle preserved in archaean rocks}}.
\newblock {\emph{\JournalTitle{Nature}}} \textbf{\bibinfo{volume}{579}},
  \bibinfo{pages}{240--244} (\bibinfo{year}{2020}).

\bibitem{elkins2008}
\bibinfo{author}{Elkins-Tanton, L.~T.}
\newblock \bibinfo{journal}{\bibinfo{title}{Linked magma ocean solidification
  and atmospheric growth for {E}arth and {M}ars}}.
\newblock {\emph{\JournalTitle{Earth and Planetary Science Letters}}}
  \textbf{\bibinfo{volume}{271}}, \bibinfo{pages}{181--191}
  (\bibinfo{year}{2008}).

\bibitem{dasgupta2019}
\bibinfo{author}{Dasgupta, R.} \& \bibinfo{author}{Grewal, D.~S.}
\newblock \bibinfo{title}{Origin and {E}arly {D}ifferentiation of {C}arbon and
  {A}ssociated {L}ife-{E}ssential {V}olatile {E}lements on {E}arth}.
\newblock In \emph{\bibinfo{booktitle}{Deep Carbon}}, \bibinfo{pages}{4--39}
  (\bibinfo{publisher}{Cambridge University Press}, \bibinfo{year}{2019}).

\bibitem{deNiem2012}
\bibinfo{author}{de~Niem, D.}, \bibinfo{author}{K{\"u}hrt, E.},
  \bibinfo{author}{Morbidelli, A.} \& \bibinfo{author}{Motschmann, U.}
\newblock \bibinfo{journal}{\bibinfo{title}{Atmospheric erosion and
  replenishment induced by impacts upon the {E}arth and {M}ars during a heavy
  bombardment}}.
\newblock {\emph{\JournalTitle{Icarus}}} \textbf{\bibinfo{volume}{221}},
  \bibinfo{pages}{495--507} (\bibinfo{year}{2012}).

\bibitem{gendaabe2005}
\bibinfo{author}{Genda, H.} \& \bibinfo{author}{Abe, Y.}
\newblock \bibinfo{journal}{\bibinfo{title}{Enhanced atmospheric loss on
  protoplanets at the giant impact phase in the presence of oceans}}.
\newblock {\emph{\JournalTitle{Nature}}} \textbf{\bibinfo{volume}{433}},
  \bibinfo{pages}{842--844} (\bibinfo{year}{2005}).

\bibitem{sakuraba2019}
\bibinfo{author}{Sakuraba, H.}, \bibinfo{author}{Kurokawa, H.} \&
  \bibinfo{author}{Genda, H.}
\newblock \bibinfo{journal}{\bibinfo{title}{Impact degassing and atmospheric
  erosion on {V}enus, {E}arth, and {M}ars during the late accretion}}.
\newblock {\emph{\JournalTitle{Icarus}}} \textbf{\bibinfo{volume}{317}},
  \bibinfo{pages}{48--58} (\bibinfo{year}{2019}).

\bibitem{bergin2015}
\bibinfo{author}{Bergin, E.~A.}, \bibinfo{author}{Blake, G.~A.},
  \bibinfo{author}{Ciesla, F.}, \bibinfo{author}{Hirschmann, M.~M.} \&
  \bibinfo{author}{Li, J.}
\newblock \bibinfo{journal}{\bibinfo{title}{Tracing the ingredients for a
  habitable earth from interstellar space through planet formation}}.
\newblock {\emph{\JournalTitle{Proceedings of the National Academy of
  Sciences}}} \textbf{\bibinfo{volume}{112}}, \bibinfo{pages}{8965--8970}
  (\bibinfo{year}{2015}).

\bibitem{grewal2019}
\bibinfo{author}{Grewal, D.~S.}, \bibinfo{author}{Dasgupta, R.},
  \bibinfo{author}{Sun, C.}, \bibinfo{author}{Tsuno, K.} \&
  \bibinfo{author}{Costin, G.}
\newblock \bibinfo{journal}{\bibinfo{title}{Delivery of carbon, nitrogen, and
  sulfur to the silicate {E}arth by a giant impact}}.
\newblock {\emph{\JournalTitle{Science advances}}}
  \textbf{\bibinfo{volume}{5}}, \bibinfo{pages}{eaau3669}
  (\bibinfo{year}{2019}).

\bibitem{hirschmann2018}
\bibinfo{author}{Hirschmann, M.~M.}
\newblock \bibinfo{journal}{\bibinfo{title}{Comparative deep {E}arth volatile
  cycles: The case for {C} recycling from exosphere/mantle fractionation of
  major ({H$_2$O}, {C}, {N}) volatiles and from {H$_2$O}/{C}e, {CO$_2$/B}a, and
  {CO$_2$/N}b exosphere ratios}}.
\newblock {\emph{\JournalTitle{Earth and Planetary Science Letters}}}
  \textbf{\bibinfo{volume}{502}}, \bibinfo{pages}{262--273}
  (\bibinfo{year}{2018}).

\bibitem{dasgupta2013}
\bibinfo{author}{Dasgupta, R.}, \bibinfo{author}{Chi, H.},
  \bibinfo{author}{Shimizu, N.}, \bibinfo{author}{Buono, A.~S.} \&
  \bibinfo{author}{Walker, D.}
\newblock \bibinfo{journal}{\bibinfo{title}{Carbon solution and partitioning
  between metallic and silicate melts in a shallow magma ocean: Implications
  for the origin and distribution of terrestrial carbon}}.
\newblock {\emph{\JournalTitle{Geochimica et Cosmochimica Acta}}}
  \textbf{\bibinfo{volume}{102}}, \bibinfo{pages}{191--212}
  (\bibinfo{year}{2013}).

\bibitem{moore1998}
\bibinfo{author}{Moore, G.}, \bibinfo{author}{Vennemann, T.} \&
  \bibinfo{author}{Carmichael, I.}
\newblock \bibinfo{journal}{\bibinfo{title}{An empirical model for the
  solubility of {H$_2$O} in magmas to 3 kilobars}}.
\newblock {\emph{\JournalTitle{american Mineralogist}}}
  \textbf{\bibinfo{volume}{83}}, \bibinfo{pages}{36--42}
  (\bibinfo{year}{1998}).

\bibitem{pan1991}
\bibinfo{author}{Pan, V.}, \bibinfo{author}{Holloway, J.~R.} \&
  \bibinfo{author}{Hervig, R.~L.}
\newblock \bibinfo{journal}{\bibinfo{title}{The pressure and temperature
  dependence of carbon dioxide solubility in tholeiitic basalt melts}}.
\newblock {\emph{\JournalTitle{Geochimica et Cosmochimica Acta}}}
  \textbf{\bibinfo{volume}{55}}, \bibinfo{pages}{1587--1595}
  (\bibinfo{year}{1991}).

\bibitem{miyazaki2004}
\bibinfo{author}{Miyazaki, A.}, \bibinfo{author}{Hiyagon, H.},
  \bibinfo{author}{Sugiura, N.}, \bibinfo{author}{Hirose, K.} \&
  \bibinfo{author}{Takahashi, E.}
\newblock \bibinfo{journal}{\bibinfo{title}{Solubilities of nitrogen and noble
  gases in silicate melts under various oxygen fugacities: implications for the
  origin and degassing history of nitrogen and noble gases in the {E}arth}}.
\newblock {\emph{\JournalTitle{Geochimica et Cosmochimica Acta}}}
  \textbf{\bibinfo{volume}{68}}, \bibinfo{pages}{387--401}
  (\bibinfo{year}{2004}).

\bibitem{dalou2017}
\bibinfo{author}{Dalou, C.}, \bibinfo{author}{Hirschmann, M.~M.},
  \bibinfo{author}{von~der Handt, A.}, \bibinfo{author}{Mosenfelder, J.} \&
  \bibinfo{author}{Armstrong, L.~S.}
\newblock \bibinfo{journal}{\bibinfo{title}{Nitrogen and carbon fractionation
  during core--mantle differentiation at shallow depth}}.
\newblock {\emph{\JournalTitle{Earth and Planetary Science Letters}}}
  \textbf{\bibinfo{volume}{458}}, \bibinfo{pages}{141--151}
  (\bibinfo{year}{2017}).

\bibitem{grewal2021}
\bibinfo{author}{Grewal, D.~S.}, \bibinfo{author}{Dasgupta, R.},
  \bibinfo{author}{Hough, T.} \& \bibinfo{author}{Farnell, A.}
\newblock \bibinfo{journal}{\bibinfo{title}{Rates of protoplanetary accretion
  and differentiation set nitrogen budget of rocky planets}}.
\newblock {\emph{\JournalTitle{Nature Geoscience}}}
  \textbf{\bibinfo{volume}{14}}, \bibinfo{pages}{369--376}
  (\bibinfo{year}{2021}).

\bibitem{canup2001}
\bibinfo{author}{Canup, R.~M.} \& \bibinfo{author}{Asphaug, E.}
\newblock \bibinfo{journal}{\bibinfo{title}{Origin of the {M}oon in a giant
  impact near the end of the {E}arth's formation}}.
\newblock {\emph{\JournalTitle{Nature}}} \textbf{\bibinfo{volume}{412}},
  \bibinfo{pages}{708--712} (\bibinfo{year}{2001}).

\bibitem{canup2004}
\bibinfo{author}{Canup, R.~M.}
\newblock \bibinfo{journal}{\bibinfo{title}{Simulations of a late lunar-forming
  impact}}.
\newblock {\emph{\JournalTitle{Icarus}}} \textbf{\bibinfo{volume}{168}},
  \bibinfo{pages}{433--456} (\bibinfo{year}{2004}).

\bibitem{salvador2017}
\bibinfo{author}{Salvador, A.} \emph{et~al.}
\newblock \bibinfo{journal}{\bibinfo{title}{The relative influence of
  {H}$_2${O} and {C}{O}$_2$ on the primitive surface conditions and evolution
  of rocky planets}}.
\newblock {\emph{\JournalTitle{Journal of Geophysical Research: Planets}}}
  (\bibinfo{year}{2017}).

\bibitem{nikolaou2019}
\bibinfo{author}{Nikolaou, A.} \emph{et~al.}
\newblock \bibinfo{journal}{\bibinfo{title}{What factors affect the duration
  and outgassing of the terrestrial magma ocean?}}
\newblock {\emph{\JournalTitle{The Astrophysical Journal}}}
  \textbf{\bibinfo{volume}{875}}, \bibinfo{pages}{11} (\bibinfo{year}{2019}).

\bibitem{hamano2013}
\bibinfo{author}{Hamano, K.}, \bibinfo{author}{Abe, Y.} \&
  \bibinfo{author}{Genda, H.}
\newblock \bibinfo{journal}{\bibinfo{title}{Emergence of two types of
  terrestrial planet on solidification of magma ocean}}.
\newblock {\emph{\JournalTitle{Nature}}} \textbf{\bibinfo{volume}{497}},
  \bibinfo{pages}{607--610} (\bibinfo{year}{2013}).

\bibitem{appel1998}
\bibinfo{author}{Appel, P.~W.}, \bibinfo{author}{Fedo, C.~M.},
  \bibinfo{author}{Moorbath, S.} \& \bibinfo{author}{Myers, J.~S.}
\newblock \bibinfo{journal}{\bibinfo{title}{Early {A}rchaean {I}sua
  supracrustal belt, {W}est {G}reenland: pilot study of the {I}sua
  multidisciplinary research project}}.
\newblock {\emph{\JournalTitle{Geology of Greenland Survey Bulletin}}}
  \textbf{\bibinfo{volume}{180}}, \bibinfo{pages}{94--99}
  (\bibinfo{year}{1998}).

\bibitem{komiya1999}
\bibinfo{author}{Komiya, T.} \emph{et~al.}
\newblock \bibinfo{journal}{\bibinfo{title}{Plate tectonics at 3.8--3.7 {G}a:
  {F}ield evidence from the {I}sua accretionary complex, southern {W}est
  {G}reenland}}.
\newblock {\emph{\JournalTitle{The Journal of Geology}}}
  \textbf{\bibinfo{volume}{107}}, \bibinfo{pages}{515--554}
  (\bibinfo{year}{1999}).

\bibitem{wilde2001}
\bibinfo{author}{Wilde, S.~A.}, \bibinfo{author}{Valley, J.~W.},
  \bibinfo{author}{Peck, W.~H.} \& \bibinfo{author}{Graham, C.~M.}
\newblock \bibinfo{journal}{\bibinfo{title}{Evidence from detrital zircons for
  the existence of continental crust and oceans on the {E}arth 4.4 {G}yr ago}}.
\newblock {\emph{\JournalTitle{Nature}}} \textbf{\bibinfo{volume}{409}},
  \bibinfo{pages}{175--178} (\bibinfo{year}{2001}).

\bibitem{hopkins2008}
\bibinfo{author}{Hopkins, M.}, \bibinfo{author}{Harrison, T.~M.} \&
  \bibinfo{author}{Manning, C.~E.}
\newblock \bibinfo{journal}{\bibinfo{title}{Low heat flow inferred from> 4
  {Gyr} zircons suggests {H}adean plate boundary interactions}}.
\newblock {\emph{\JournalTitle{Nature}}} \textbf{\bibinfo{volume}{456}},
  \bibinfo{pages}{493--496} (\bibinfo{year}{2008}).

\bibitem{korenaga2013}
\bibinfo{author}{Korenaga, J.}
\newblock \bibinfo{journal}{\bibinfo{title}{Initiation and evolution of plate
  tectonics on {E}arth: theories and observations}}.
\newblock {\emph{\JournalTitle{Annual review of earth and planetary sciences}}}
  \textbf{\bibinfo{volume}{41}}, \bibinfo{pages}{117--151}
  (\bibinfo{year}{2013}).

\bibitem{foley2019}
\bibinfo{author}{Foley, B.~J.}
\newblock \bibinfo{journal}{\bibinfo{title}{Habitability of {E}arth-like
  stagnant lid planets: {C}limate evolution and recovery from snowball
  states}}.
\newblock {\emph{\JournalTitle{The Astrophysical Journal}}}
  \textbf{\bibinfo{volume}{875}}, \bibinfo{pages}{72} (\bibinfo{year}{2019}).

\bibitem{krissansen2018}
\bibinfo{author}{Krissansen-Totton, J.}, \bibinfo{author}{Arney, G.~N.} \&
  \bibinfo{author}{Catling, D.~C.}
\newblock \bibinfo{journal}{\bibinfo{title}{Constraining the climate and ocean
  ph of the early {E}arth with a geological carbon cycle model}}.
\newblock {\emph{\JournalTitle{Proceedings of the National Academy of
  Sciences}}} \bibinfo{pages}{201721296} (\bibinfo{year}{2018}).

\bibitem{pierrehumbert2010}
\bibinfo{author}{Pierrehumbert, R.~T.}
\newblock \emph{\bibinfo{title}{Principles of planetary climate}}
  (\bibinfo{publisher}{Cambridge University Press}, \bibinfo{year}{2010}).

\bibitem{tucker2014}
\bibinfo{author}{Tucker, J.~M.} \& \bibinfo{author}{Mukhopadhyay, S.}
\newblock \bibinfo{journal}{\bibinfo{title}{Evidence for multiple magma ocean
  outgassing and atmospheric loss episodes from mantle noble gases}}.
\newblock {\emph{\JournalTitle{Earth and Planetary Science Letters}}}
  \textbf{\bibinfo{volume}{393}}, \bibinfo{pages}{254--265}
  (\bibinfo{year}{2014}).

\bibitem{hu2019}
\bibinfo{author}{Hu, R.} \& \bibinfo{author}{Diaz, H.~D.}
\newblock \bibinfo{journal}{\bibinfo{title}{Stability of nitrogen in planetary
  atmospheres in contact with liquid water}}.
\newblock {\emph{\JournalTitle{The Astrophysical Journal}}}
  \textbf{\bibinfo{volume}{886}}, \bibinfo{pages}{126} (\bibinfo{year}{2019}).

\bibitem{stueken2016}
\bibinfo{author}{St{\"u}eken, E.~E.}, \bibinfo{author}{Kipp, M.~A.},
  \bibinfo{author}{Koehler, M.~C.} \& \bibinfo{author}{Buick, R.}
\newblock \bibinfo{journal}{\bibinfo{title}{The evolution of {E}arth's
  biogeochemical nitrogen cycle}}.
\newblock {\emph{\JournalTitle{Earth-Science Reviews}}}
  \textbf{\bibinfo{volume}{160}}, \bibinfo{pages}{220--239}
  (\bibinfo{year}{2016}).

\bibitem{catling2020}
\bibinfo{author}{Catling, D.~C.} \& \bibinfo{author}{Zahnle, K.~J.}
\newblock \bibinfo{journal}{\bibinfo{title}{The {A}rchean atmosphere}}.
\newblock {\emph{\JournalTitle{Science Advances}}}
  \textbf{\bibinfo{volume}{6}}, \bibinfo{pages}{eaax1420}
  (\bibinfo{year}{2020}).

\bibitem{morbidelli2018}
\bibinfo{author}{Morbidelli, A.} \emph{et~al.}
\newblock \bibinfo{journal}{\bibinfo{title}{The timeline of the lunar
  bombardment: Revisited}}.
\newblock {\emph{\JournalTitle{Icarus}}} \textbf{\bibinfo{volume}{305}},
  \bibinfo{pages}{262--276} (\bibinfo{year}{2018}).

\bibitem{schlichting2015}
\bibinfo{author}{Schlichting, H.~E.}, \bibinfo{author}{Sari, R.} \&
  \bibinfo{author}{Yalinewich, A.}
\newblock \bibinfo{journal}{\bibinfo{title}{Atmospheric mass loss during planet
  formation: The importance of planetesimal impacts}}.
\newblock {\emph{\JournalTitle{Icarus}}} \textbf{\bibinfo{volume}{247}},
  \bibinfo{pages}{81--94} (\bibinfo{year}{2015}).

\bibitem{goldreich2004}
\bibinfo{author}{Goldreich, P.}, \bibinfo{author}{Lithwick, Y.} \&
  \bibinfo{author}{Sari, R.}
\newblock \bibinfo{journal}{\bibinfo{title}{Final stages of planet formation}}.
\newblock {\emph{\JournalTitle{The Astrophysical Journal}}}
  \textbf{\bibinfo{volume}{614}}, \bibinfo{pages}{497} (\bibinfo{year}{2004}).

\bibitem{kenyon2006}
\bibinfo{author}{Kenyon, S.~J.} \& \bibinfo{author}{Bromley, B.~C.}
\newblock \bibinfo{journal}{\bibinfo{title}{Terrestrial planet formation. {I}.
  {T}he transition from oligarchic growth to chaotic growth}}.
\newblock {\emph{\JournalTitle{The Astronomical Journal}}}
  \textbf{\bibinfo{volume}{131}}, \bibinfo{pages}{1837} (\bibinfo{year}{2006}).

\bibitem{kokubo2010}
\bibinfo{author}{Kokubo, E.} \& \bibinfo{author}{Genda, H.}
\newblock \bibinfo{journal}{\bibinfo{title}{Formation of terrestrial planets
  from protoplanets under a realistic accretion condition}}.
\newblock {\emph{\JournalTitle{The Astrophysical Journal Letters}}}
  \textbf{\bibinfo{volume}{714}}, \bibinfo{pages}{L21} (\bibinfo{year}{2010}).

\bibitem{bottke2005}
\bibinfo{author}{Bottke, W.~F.} \emph{et~al.}
\newblock \bibinfo{journal}{\bibinfo{title}{The fossilized size distribution of
  the main asteroid belt}}.
\newblock {\emph{\JournalTitle{Icarus}}} \textbf{\bibinfo{volume}{175}},
  \bibinfo{pages}{111--140} (\bibinfo{year}{2005}).

\bibitem{bottke2015}
\bibinfo{author}{Bottke, W.} \emph{et~al.}
\newblock \bibinfo{journal}{\bibinfo{title}{Dating the {M}oon-forming impact
  event with asteroidal meteorites}}.
\newblock {\emph{\JournalTitle{Science}}} \textbf{\bibinfo{volume}{348}},
  \bibinfo{pages}{321--323} (\bibinfo{year}{2015}).

\bibitem{bottke2010science}
\bibinfo{author}{Bottke, W.~F.}, \bibinfo{author}{Walker, R.~J.},
  \bibinfo{author}{Day, J.~M.}, \bibinfo{author}{Nesvorny, D.} \&
  \bibinfo{author}{Elkins-Tanton, L.}
\newblock \bibinfo{journal}{\bibinfo{title}{Stochastic late accretion to
  {E}arth, the {M}oon, and {M}ars}}.
\newblock {\emph{\JournalTitle{Science}}} \textbf{\bibinfo{volume}{330}},
  \bibinfo{pages}{1527--1530} (\bibinfo{year}{2010}).

\bibitem{levison2015}
\bibinfo{author}{Levison, H.~F.}, \bibinfo{author}{Kretke, K.~A.},
  \bibinfo{author}{Walsh, K.~J.} \& \bibinfo{author}{Bottke, W.~F.}
\newblock \bibinfo{journal}{\bibinfo{title}{Growing the terrestrial planets
  from the gradual accumulation of submeter-sized objects}}.
\newblock {\emph{\JournalTitle{Proceedings of the National Academy of
  Sciences}}} \textbf{\bibinfo{volume}{112}}, \bibinfo{pages}{14180--14185}
  (\bibinfo{year}{2015}).

\bibitem{brasser2016}
\bibinfo{author}{Brasser, R.}, \bibinfo{author}{Mojzsis, S.},
  \bibinfo{author}{Werner, S.}, \bibinfo{author}{Matsumura, S.} \&
  \bibinfo{author}{Ida, S.}
\newblock \bibinfo{journal}{\bibinfo{title}{Late veneer and late accretion to
  the terrestrial planets}}.
\newblock {\emph{\JournalTitle{Earth and Planetary Science Letters}}}
  \textbf{\bibinfo{volume}{455}}, \bibinfo{pages}{85--93}
  (\bibinfo{year}{2016}).

\bibitem{albarede2009}
\bibinfo{author}{Albarede, F.}
\newblock \bibinfo{journal}{\bibinfo{title}{Volatile accretion history of the
  terrestrial planets and dynamic implications}}.
\newblock {\emph{\JournalTitle{Nature}}} \textbf{\bibinfo{volume}{461}},
  \bibinfo{pages}{1227--1233} (\bibinfo{year}{2009}).

\bibitem{chou1978}
\bibinfo{author}{Chou, C.~L.}
\newblock \bibinfo{journal}{\bibinfo{title}{Fractionation of siderophile
  elements in the {E}arth's upper mantle.}}
\newblock {\emph{\JournalTitle{Lunar and Planetary Sciences}}}
  \textbf{\bibinfo{volume}{IX}}, \bibinfo{pages}{219--230}
  (\bibinfo{year}{1978}).

\bibitem{wanke1984}
\bibinfo{author}{W{\"a}nke, H.}, \bibinfo{author}{Dreibus, G.} \&
  \bibinfo{author}{Jagoutz, E.}
\newblock \bibinfo{title}{Mantle chemistry and accretion history of the
  {E}arth}.
\newblock In \emph{\bibinfo{booktitle}{Archaean geochemistry}},
  \bibinfo{pages}{1--24} (\bibinfo{publisher}{Springer}, \bibinfo{year}{1984}).

\bibitem{wade2005}
\bibinfo{author}{Wade, J.} \& \bibinfo{author}{Wood, B.}
\newblock \bibinfo{journal}{\bibinfo{title}{Core formation and the oxidation
  state of the {E}arth}}.
\newblock {\emph{\JournalTitle{Earth and Planetary Science Letters}}}
  \textbf{\bibinfo{volume}{236}}, \bibinfo{pages}{78--95}
  (\bibinfo{year}{2005}).

\bibitem{schonbachler2010}
\bibinfo{author}{Sch{\"o}nb{\"a}chler, M.}, \bibinfo{author}{Carlson, R.},
  \bibinfo{author}{Horan, M.}, \bibinfo{author}{Mock, T.} \&
  \bibinfo{author}{Hauri, E.}
\newblock \bibinfo{journal}{\bibinfo{title}{Heterogeneous accretion and the
  moderately volatile element budget of {E}arth}}.
\newblock {\emph{\JournalTitle{Science}}} \textbf{\bibinfo{volume}{328}},
  \bibinfo{pages}{884--887} (\bibinfo{year}{2010}).

\bibitem{marchi2018}
\bibinfo{author}{Marchi, S.}, \bibinfo{author}{Canup, R.} \&
  \bibinfo{author}{Walker, R.}
\newblock \bibinfo{journal}{\bibinfo{title}{Heterogeneous delivery of silicate
  and metal to the {E}arth by large planetesimals}}.
\newblock {\emph{\JournalTitle{Nature Geoscience}}}
  \textbf{\bibinfo{volume}{11}}, \bibinfo{pages}{77} (\bibinfo{year}{2018}).

\bibitem{korenaga2017}
\bibinfo{author}{Korenaga, J.}, \bibinfo{author}{Planavsky, N.~J.} \&
  \bibinfo{author}{Evans, D.~A.}
\newblock \bibinfo{journal}{\bibinfo{title}{Global water cycle and the
  coevolution of the {E}arth's interior and surface environment}}.
\newblock {\emph{\JournalTitle{Phil. Trans. R. Soc. A}}}
  \textbf{\bibinfo{volume}{375}}, \bibinfo{pages}{20150393}
  (\bibinfo{year}{2017}).

\bibitem{peslier2017}
\bibinfo{author}{Peslier, A.~H.}, \bibinfo{author}{Sch{\"o}nb{\"a}chler, M.},
  \bibinfo{author}{Busemann, H.} \& \bibinfo{author}{Karato, S.-I.}
\newblock \bibinfo{journal}{\bibinfo{title}{Water in the {E}arth's interior:
  distribution and origin}}.
\newblock {\emph{\JournalTitle{Space Science Reviews}}}
  \textbf{\bibinfo{volume}{212}}, \bibinfo{pages}{743--810}
  (\bibinfo{year}{2017}).

\bibitem{tsuno2018}
\bibinfo{author}{Tsuno, K.}, \bibinfo{author}{Grewal, D.~S.} \&
  \bibinfo{author}{Dasgupta, R.}
\newblock \bibinfo{journal}{\bibinfo{title}{Core-mantle fractionation of carbon
  in {E}arth and {M}ars: {T}he effects of sulfur}}.
\newblock {\emph{\JournalTitle{Geochimica et Cosmochimica Acta}}}
  \textbf{\bibinfo{volume}{238}}, \bibinfo{pages}{477--495}
  (\bibinfo{year}{2018}).

\bibitem{bureau1998}
\bibinfo{author}{Bureau, H.}, \bibinfo{author}{Pineau, F.},
  \bibinfo{author}{M{\'e}trich, N.}, \bibinfo{author}{Semet, M.} \&
  \bibinfo{author}{Javoy, M.}
\newblock \bibinfo{journal}{\bibinfo{title}{A melt and fluid inclusion study of
  the gas phase at {P}iton de la {F}ournaise volcano ({R{\'e}}union
  {I}sland)}}.
\newblock {\emph{\JournalTitle{Chemical geology}}}
  \textbf{\bibinfo{volume}{147}}, \bibinfo{pages}{115--130}
  (\bibinfo{year}{1998}).

\bibitem{clesi2018}
\bibinfo{author}{Clesi, V.} \emph{et~al.}
\newblock \bibinfo{journal}{\bibinfo{title}{Low hydrogen contents in the cores
  of terrestrial planets}}.
\newblock {\emph{\JournalTitle{Science advances}}}
  \textbf{\bibinfo{volume}{4}}, \bibinfo{pages}{e1701876}
  (\bibinfo{year}{2018}).

\bibitem{malavergne2019}
\bibinfo{author}{Malavergne, V.} \emph{et~al.}
\newblock \bibinfo{journal}{\bibinfo{title}{Experimental constraints on the
  fate of {H} and {C} during planetary core-mantle differentiation.
  {I}mplications for the {E}arth}}.
\newblock {\emph{\JournalTitle{Icarus}}} \textbf{\bibinfo{volume}{321}},
  \bibinfo{pages}{473--485} (\bibinfo{year}{2019}).

\bibitem{siebert2013}
\bibinfo{author}{Siebert, J.}, \bibinfo{author}{Badro, J.},
  \bibinfo{author}{Antonangeli, D.} \& \bibinfo{author}{Ryerson, F.~J.}
\newblock \bibinfo{journal}{\bibinfo{title}{Terrestrial accretion under
  oxidizing conditions}}.
\newblock {\emph{\JournalTitle{Science}}} \textbf{\bibinfo{volume}{339}},
  \bibinfo{pages}{1194--1197} (\bibinfo{year}{2013}).

\bibitem{badro2015}
\bibinfo{author}{Badro, J.}, \bibinfo{author}{Brodholt, J.~P.},
  \bibinfo{author}{Piet, H.}, \bibinfo{author}{Siebert, J.} \&
  \bibinfo{author}{Ryerson, F.~J.}
\newblock \bibinfo{journal}{\bibinfo{title}{Core formation and core composition
  from coupled geochemical and geophysical constraints}}.
\newblock {\emph{\JournalTitle{Proceedings of the National Academy of
  Sciences}}} \textbf{\bibinfo{volume}{112}}, \bibinfo{pages}{12310--12314}
  (\bibinfo{year}{2015}).

\bibitem{armstrong2019}
\bibinfo{author}{Armstrong, K.}, \bibinfo{author}{Frost, D.~J.},
  \bibinfo{author}{McCammon, C.~A.}, \bibinfo{author}{Rubie, D.~C.} \&
  \bibinfo{author}{Ballaran, T.~B.}
\newblock \bibinfo{journal}{\bibinfo{title}{Deep magma ocean formation set the
  oxidation state of {E}arth's mantle}}.
\newblock {\emph{\JournalTitle{Science}}} \textbf{\bibinfo{volume}{365}},
  \bibinfo{pages}{903--906} (\bibinfo{year}{2019}).

\bibitem{wood2006}
\bibinfo{author}{Wood, B.~J.}, \bibinfo{author}{Walter, M.~J.} \&
  \bibinfo{author}{Wade, J.}
\newblock \bibinfo{journal}{\bibinfo{title}{Accretion of the {E}arth and
  segregation of its core}}.
\newblock {\emph{\JournalTitle{Nature}}} \textbf{\bibinfo{volume}{441}},
  \bibinfo{pages}{825} (\bibinfo{year}{2006}).

\bibitem{hirschmann2012}
\bibinfo{author}{Hirschmann, M.}, \bibinfo{author}{Withers, A.},
  \bibinfo{author}{Ardia, P.} \& \bibinfo{author}{Foley, N.}
\newblock \bibinfo{journal}{\bibinfo{title}{Solubility of molecular hydrogen in
  silicate melts and consequences for volatile evolution of terrestrial
  planets}}.
\newblock {\emph{\JournalTitle{Earth and Planetary Science Letters}}}
  \textbf{\bibinfo{volume}{345}}, \bibinfo{pages}{38--48}
  (\bibinfo{year}{2012}).

\bibitem{fischer2015}
\bibinfo{author}{Fischer, R.~A.} \emph{et~al.}
\newblock \bibinfo{journal}{\bibinfo{title}{High pressure metal--silicate
  partitioning of {N}i, {C}o, {V}, {C}r, {S}i, and {O}}}.
\newblock {\emph{\JournalTitle{Geochimica et Cosmochimica Acta}}}
  \textbf{\bibinfo{volume}{167}}, \bibinfo{pages}{177--194}
  (\bibinfo{year}{2015}).

\bibitem{rubie2015}
\bibinfo{author}{Rubie, D.~C.} \emph{et~al.}
\newblock \bibinfo{journal}{\bibinfo{title}{Accretion and differentiation of
  the terrestrial planets with implications for the compositions of
  early-formed {S}olar {S}ystem bodies and accretion of water}}.
\newblock {\emph{\JournalTitle{Icarus}}} \textbf{\bibinfo{volume}{248}},
  \bibinfo{pages}{89--108} (\bibinfo{year}{2015}).

\bibitem{hirose2013}
\bibinfo{author}{Hirose, K.}, \bibinfo{author}{Labrosse, S.} \&
  \bibinfo{author}{Hernlund, J.}
\newblock \bibinfo{journal}{\bibinfo{title}{Composition and state of the
  core}}.
\newblock {\emph{\JournalTitle{Annual Review of Earth and Planetary Sciences}}}
  \textbf{\bibinfo{volume}{41}}, \bibinfo{pages}{657--691}
  (\bibinfo{year}{2013}).

\bibitem{dauphas2014}
\bibinfo{author}{Dauphas, N.} \& \bibinfo{author}{Morbidelli, A.}
\newblock \bibinfo{title}{6.1 - {G}eochemical and {P}lanetary {D}ynamical
  {V}iews on the origin of {E}arth's {A}tmosphere and {O}ceans}.
\newblock In \bibinfo{editor}{Holland, H.~D.} \& \bibinfo{editor}{Turekian,
  K.~K.} (eds.) \emph{\bibinfo{booktitle}{Treatise on Geochemistry (Second
  Edition)}}, \bibinfo{pages}{1--35} (\bibinfo{publisher}{Elsevier},
  \bibinfo{address}{Oxford}, \bibinfo{year}{2014}), \bibinfo{edition}{second
  edition} edn.

\bibitem{forget2013}
\bibinfo{author}{Forget, F.} \emph{et~al.}
\newblock \bibinfo{journal}{\bibinfo{title}{3d modelling of the early {M}artian
  climate under a denser {CO$_2$} atmosphere: Temperatures and {CO$_2$} ice
  clouds}}.
\newblock {\emph{\JournalTitle{Icarus}}} \textbf{\bibinfo{volume}{222}},
  \bibinfo{pages}{81--99} (\bibinfo{year}{2013}).

\bibitem{wadhwa2001}
\bibinfo{author}{Wadhwa, M.}
\newblock \bibinfo{journal}{\bibinfo{title}{Redox state of {Mars'} upper mantle
  and crust from {E}u anomalies in shergottite pyroxenes}}.
\newblock {\emph{\JournalTitle{Science}}} \textbf{\bibinfo{volume}{291}},
  \bibinfo{pages}{1527--1530} (\bibinfo{year}{2001}).

\bibitem{herd2002}
\bibinfo{author}{Herd, C.~D.}, \bibinfo{author}{Borg, L.~E.},
  \bibinfo{author}{Jones, J.~H.} \& \bibinfo{author}{Papike, J.~J.}
\newblock \bibinfo{journal}{\bibinfo{title}{Oxygen fugacity and geochemical
  variations in the martian basalts: {I}mplications for martian basalt
  petrogenesis and the oxidation state of the upper mantle of {M}ars}}.
\newblock {\emph{\JournalTitle{Geochimica et Cosmochimica Acta}}}
  \textbf{\bibinfo{volume}{66}}, \bibinfo{pages}{2025--2036}
  (\bibinfo{year}{2002}).

\bibitem{hirschmann2008}
\bibinfo{author}{Hirschmann, M.~M.} \& \bibinfo{author}{Withers, A.~C.}
\newblock \bibinfo{journal}{\bibinfo{title}{Ventilation of {CO$_2$} from a
  reduced mantle and consequences for the early {M}artian greenhouse}}.
\newblock {\emph{\JournalTitle{Earth and Planetary Science Letters}}}
  \textbf{\bibinfo{volume}{270}}, \bibinfo{pages}{147--155}
  (\bibinfo{year}{2008}).

\bibitem{zhang2017}
\bibinfo{author}{Zhang, H.}, \bibinfo{author}{Hirschmann, M.},
  \bibinfo{author}{Cottrell, E.} \& \bibinfo{author}{Withers, A.}
\newblock \bibinfo{journal}{\bibinfo{title}{Effect of pressure on
  {Fe$^{3+}$/$\Sigma$Fe} ratio in a mafic magma and consequences for magma
  ocean redox gradients}}.
\newblock {\emph{\JournalTitle{Geochimica et Cosmochimica Acta}}}
  \textbf{\bibinfo{volume}{204}}, \bibinfo{pages}{83--103}
  (\bibinfo{year}{2017}).

\bibitem{seager2007}
\bibinfo{author}{Seager, S.}, \bibinfo{author}{Kuchner, M.},
  \bibinfo{author}{Hier-Majumder, C.} \& \bibinfo{author}{Militzer, B.}
\newblock \bibinfo{journal}{\bibinfo{title}{Mass-radius relationships for solid
  exoplanets}}.
\newblock {\emph{\JournalTitle{The Astrophysical Journal}}}
  \textbf{\bibinfo{volume}{669}}, \bibinfo{pages}{1279} (\bibinfo{year}{2007}).

\bibitem{idamakino1992}
\bibinfo{author}{Ida, S.} \& \bibinfo{author}{Makino, J.}
\newblock \bibinfo{journal}{\bibinfo{title}{N-body simulation of gravitational
  interaction between planetesimals and a protoplanet: I. velocity distribution
  of planetesimals}}.
\newblock {\emph{\JournalTitle{Icarus}}} \textbf{\bibinfo{volume}{96}},
  \bibinfo{pages}{107--120} (\bibinfo{year}{1992}).

\bibitem{kerridge1985}
\bibinfo{author}{Kerridge, J.~F.}
\newblock \bibinfo{journal}{\bibinfo{title}{Carbon, hydrogen and nitrogen in
  carbonaceous chondrites: {A}bundances and isotopic compositions in bulk
  samples}}.
\newblock {\emph{\JournalTitle{Geochimica et Cosmochimica Acta}}}
  \textbf{\bibinfo{volume}{49}}, \bibinfo{pages}{1707--1714}
  (\bibinfo{year}{1985}).

\bibitem{pepin1991}
\bibinfo{author}{Pepin, R.~O.}
\newblock \bibinfo{journal}{\bibinfo{title}{On the origin and early evolution
  of terrestrial planet atmospheres and meteoritic volatiles}}.
\newblock {\emph{\JournalTitle{Icarus}}} \textbf{\bibinfo{volume}{92}},
  \bibinfo{pages}{2--79} (\bibinfo{year}{1991}).

\bibitem{gradywright2003}
\bibinfo{author}{Grady, M.~M.} \& \bibinfo{author}{Wright, I.~P.}
\newblock \bibinfo{journal}{\bibinfo{title}{Elemental and isotopic abundances
  of carbon and nitrogen in meteorites}}.
\newblock {\emph{\JournalTitle{Space Science Reviews}}}
  \textbf{\bibinfo{volume}{106}}, \bibinfo{pages}{231--248}
  (\bibinfo{year}{2003}).

\bibitem{vacher2020}
\bibinfo{author}{Vacher, L.~G.} \emph{et~al.}
\newblock \bibinfo{journal}{\bibinfo{title}{Hydrogen in chondrites: {I}nfluence
  of parent body alteration and atmospheric contamination on primordial
  components}}.
\newblock {\emph{\JournalTitle{Geochimica et Cosmochimica Acta}}}
  \textbf{\bibinfo{volume}{281}}, \bibinfo{pages}{53--66}
  (\bibinfo{year}{2020}).

\bibitem{robert2003}
\bibinfo{author}{Robert, F.}
\newblock \bibinfo{journal}{\bibinfo{title}{The {D/H} ratio in chondrites}}.
\newblock {\emph{\JournalTitle{Space Science Reviews}}}
  \textbf{\bibinfo{volume}{106}}, \bibinfo{pages}{87--101}
  (\bibinfo{year}{2003}).

\bibitem{piani2020}
\bibinfo{author}{Piani, L.} \emph{et~al.}
\newblock \bibinfo{journal}{\bibinfo{title}{Earth's water may have been
  inherited from material similar to enstatite chondrite meteorites}}.
\newblock {\emph{\JournalTitle{Science}}} \textbf{\bibinfo{volume}{369}},
  \bibinfo{pages}{1110--1113} (\bibinfo{year}{2020}).

\bibitem{shuvalov2009}
\bibinfo{author}{Shuvalov, V.}
\newblock \bibinfo{journal}{\bibinfo{title}{Atmospheric erosion induced by
  oblique impacts}}.
\newblock {\emph{\JournalTitle{Meteoritics \& Planetary Science}}}
  \textbf{\bibinfo{volume}{44}}, \bibinfo{pages}{1095--1105}
  (\bibinfo{year}{2009}).

\bibitem{svetsov2007}
\bibinfo{author}{Svetsov, V.}
\newblock \bibinfo{journal}{\bibinfo{title}{Atmospheric erosion and
  replenishment induced by impacts of cosmic bodies upon the {E}arth and
  {M}ars}}.
\newblock {\emph{\JournalTitle{Solar System Research}}}
  \textbf{\bibinfo{volume}{41}}, \bibinfo{pages}{28--41}
  (\bibinfo{year}{2007}).

\bibitem{svetsov2000}
\bibinfo{author}{Svetsov, V.}
\newblock \bibinfo{journal}{\bibinfo{title}{On the efficiency of the impact
  mechanism of atmospheric erosion}}.
\newblock {\emph{\JournalTitle{Solar Syst. Res.}}}
  \textbf{\bibinfo{volume}{34}}, \bibinfo{pages}{398--410}
  (\bibinfo{year}{2000}).

\bibitem{hosono2019}
\bibinfo{author}{Hosono, N.}, \bibinfo{author}{Karato, S.-i.},
  \bibinfo{author}{Makino, J.} \& \bibinfo{author}{Saitoh, T.~R.}
\newblock \bibinfo{journal}{\bibinfo{title}{Terrestrial magma ocean origin of
  the {M}oon}}.
\newblock {\emph{\JournalTitle{Nature Geoscience}}} \bibinfo{pages}{1}
  (\bibinfo{year}{2019}).

\bibitem{kaib2015}
\bibinfo{author}{Kaib, N.~A.} \& \bibinfo{author}{Cowan, N.~B.}
\newblock \bibinfo{journal}{\bibinfo{title}{The feeding zones of terrestrial
  planets and insights into {M}oon formation}}.
\newblock {\emph{\JournalTitle{Icarus}}} \textbf{\bibinfo{volume}{252}},
  \bibinfo{pages}{161--174} (\bibinfo{year}{2015}).

\bibitem{kegerreis2019}
\bibinfo{author}{Kegerreis, J.} \emph{et~al.}
\newblock \bibinfo{journal}{\bibinfo{title}{Planetary giant impacts:
  convergence of high-resolution simulations using efficient spherical initial
  conditions and swift}}.
\newblock {\emph{\JournalTitle{Monthly Notices of the Royal Astronomical
  Society}}} \textbf{\bibinfo{volume}{487}}, \bibinfo{pages}{5029--5040}
  (\bibinfo{year}{2019}).

\bibitem{kegerreis2020}
\bibinfo{author}{Kegerreis, J.~A.} \emph{et~al.}
\newblock \bibinfo{journal}{\bibinfo{title}{Atmospheric erosion by giant
  impacts onto terrestrial planets: A scaling law for any speed, angle, mass,
  and density}}.
\newblock {\emph{\JournalTitle{The Astrophysical Journal Letters}}}
  \textbf{\bibinfo{volume}{901}}, \bibinfo{pages}{L31} (\bibinfo{year}{2020}).

\bibitem{rubie2003}
\bibinfo{author}{Rubie, D.}, \bibinfo{author}{Melosh, H.},
  \bibinfo{author}{Reid, J.}, \bibinfo{author}{Liebske, C.} \&
  \bibinfo{author}{Righter, K.}
\newblock \bibinfo{journal}{\bibinfo{title}{Mechanisms of metal--silicate
  equilibration in the terrestrial magma ocean}}.
\newblock {\emph{\JournalTitle{Earth and Planetary Science Letters}}}
  \textbf{\bibinfo{volume}{205}}, \bibinfo{pages}{239--255}
  (\bibinfo{year}{2003}).

\bibitem{rubie2015treatise}
\bibinfo{author}{Rubie, D.}, \bibinfo{author}{Nimmo, F.} \&
  \bibinfo{author}{Melosh, H.}
\newblock \emph{\bibinfo{title}{Formation of {E}arth's {C}ore. {T}reatise on
  {G}eophysics. {V}. 1/{E}d. in {C}hief {G}. {S}chubert}}
  (\bibinfo{publisher}{Elsevier B.V.}, \bibinfo{year}{2015}).

\bibitem{kleine2002}
\bibinfo{author}{Kleine, T.}, \bibinfo{author}{M{\"u}nker, C.},
  \bibinfo{author}{Mezger, K.} \& \bibinfo{author}{Palme, H.}
\newblock \bibinfo{journal}{\bibinfo{title}{Rapid accretion and early core
  formation on asteroids and the terrestrial planets from hf--w chronometry}}.
\newblock {\emph{\JournalTitle{Nature}}} \textbf{\bibinfo{volume}{418}},
  \bibinfo{pages}{952} (\bibinfo{year}{2002}).

\bibitem{pritchard2000}
\bibinfo{author}{Pritchard, M.} \& \bibinfo{author}{Stevenson, D.}
\newblock \bibinfo{journal}{\bibinfo{title}{Thermal aspects of a lunar origin
  by giant impact}}.
\newblock {\emph{\JournalTitle{Origin of the {E}arth and {M}oon}}}
  \textbf{\bibinfo{volume}{1}}, \bibinfo{pages}{179--196}
  (\bibinfo{year}{2000}).

\bibitem{deguen2011}
\bibinfo{author}{Deguen, R.}, \bibinfo{author}{Olson, P.} \&
  \bibinfo{author}{Cardin, P.}
\newblock \bibinfo{journal}{\bibinfo{title}{Experiments on turbulent
  metal-silicate mixing in a magma ocean}}.
\newblock {\emph{\JournalTitle{Earth and Planetary Science Letters}}}
  \textbf{\bibinfo{volume}{310}}, \bibinfo{pages}{303--313}
  (\bibinfo{year}{2011}).

\bibitem{hier2017}
\bibinfo{author}{Hier-Majumder, S.} \& \bibinfo{author}{Hirschmann, M.~M.}
\newblock \bibinfo{journal}{\bibinfo{title}{The origin of volatiles in the
  {E}arth's mantle}}.
\newblock {\emph{\JournalTitle{Geochemistry, Geophysics, Geosystems}}}
  \textbf{\bibinfo{volume}{18}}, \bibinfo{pages}{3078--3092}
  (\bibinfo{year}{2017}).

\bibitem{li2013}
\bibinfo{author}{Li, Y.}, \bibinfo{author}{Wiedenbeck, M.},
  \bibinfo{author}{Shcheka, S.} \& \bibinfo{author}{Keppler, H.}
\newblock \bibinfo{journal}{\bibinfo{title}{Nitrogen solubility in upper mantle
  minerals}}.
\newblock {\emph{\JournalTitle{Earth and Planetary Science Letters}}}
  \textbf{\bibinfo{volume}{377}}, \bibinfo{pages}{311--323}
  (\bibinfo{year}{2013}).

\bibitem{palme2014}
\bibinfo{author}{Palme, H.} \& \bibinfo{author}{O’Neill, H.}
\newblock \bibinfo{title}{Cosmochemical {E}stimates of {M}antle {C}omposition.
  {P}lanets, {A}steriods, {C}omets and {T}he {S}olar {S}ystem, {V}olume 2 of
  {T}reatise on {G}eochemistry. {E}dited by {A}ndrew {M}. {D}avis}
  (\bibinfo{year}{2014}).

\bibitem{chi2014}
\bibinfo{author}{Chi, H.}, \bibinfo{author}{Dasgupta, R.},
  \bibinfo{author}{Duncan, M.~S.} \& \bibinfo{author}{Shimizu, N.}
\newblock \bibinfo{journal}{\bibinfo{title}{Partitioning of carbon between
  {Fe}-rich alloy melt and silicate melt in a magma ocean--implications for the
  abundance and origin of volatiles in {E}arth, {M}ars, and the {M}oon}}.
\newblock {\emph{\JournalTitle{Geochimica et Cosmochimica Acta}}}
  \textbf{\bibinfo{volume}{139}}, \bibinfo{pages}{447--471}
  (\bibinfo{year}{2014}).

\bibitem{armstrong2015}
\bibinfo{author}{Armstrong, L.~S.}, \bibinfo{author}{Hirschmann, M.~M.},
  \bibinfo{author}{Stanley, B.~D.}, \bibinfo{author}{Falksen, E.~G.} \&
  \bibinfo{author}{Jacobsen, S.~D.}
\newblock \bibinfo{journal}{\bibinfo{title}{Speciation and solubility of
  reduced {C--O--H--N} volatiles in mafic melt: {I}mplications for volcanism,
  atmospheric evolution, and deep volatile cycles in the terrestrial planets}}.
\newblock {\emph{\JournalTitle{Geochimica et Cosmochimica Acta}}}
  \textbf{\bibinfo{volume}{171}}, \bibinfo{pages}{283--302}
  (\bibinfo{year}{2015}).

\bibitem{okuchi1997}
\bibinfo{author}{Okuchi, T.}
\newblock \bibinfo{journal}{\bibinfo{title}{Hydrogen partitioning into molten
  iron at high pressure: implications for {E}arth's core}}.
\newblock {\emph{\JournalTitle{Science}}} \textbf{\bibinfo{volume}{278}},
  \bibinfo{pages}{1781--1784} (\bibinfo{year}{1997}).

\bibitem{roskosz2013}
\bibinfo{author}{Roskosz, M.}, \bibinfo{author}{Bouhifd, M.~A.},
  \bibinfo{author}{Jephcoat, A.}, \bibinfo{author}{Marty, B.} \&
  \bibinfo{author}{Mysen, B.}
\newblock \bibinfo{journal}{\bibinfo{title}{Nitrogen solubility in molten metal
  and silicate at high pressure and temperature}}.
\newblock {\emph{\JournalTitle{Geochimica et Cosmochimica Acta}}}
  \textbf{\bibinfo{volume}{121}}, \bibinfo{pages}{15--28}
  (\bibinfo{year}{2013}).

\bibitem{li2016carbon}
\bibinfo{author}{Li, Y.}, \bibinfo{author}{Dasgupta, R.},
  \bibinfo{author}{Tsuno, K.}, \bibinfo{author}{Monteleone, B.} \&
  \bibinfo{author}{Shimizu, N.}
\newblock \bibinfo{journal}{\bibinfo{title}{Carbon and sulfur budget of the
  silicate {E}arth explained by accretion of differentiated planetary
  embryos}}.
\newblock {\emph{\JournalTitle{Nature Geoscience}}}
  \textbf{\bibinfo{volume}{9}}, \bibinfo{pages}{781--785}
  (\bibinfo{year}{2016}).

\bibitem{fischer2020}
\bibinfo{author}{Fischer, R.~A.}, \bibinfo{author}{Cottrell, E.},
  \bibinfo{author}{Hauri, E.}, \bibinfo{author}{Lee, K.~K.} \&
  \bibinfo{author}{Le~Voyer, M.}
\newblock \bibinfo{journal}{\bibinfo{title}{The carbon content of {E}arth and
  its core}}.
\newblock {\emph{\JournalTitle{Proceedings of the National Academy of
  Sciences}}} \textbf{\bibinfo{volume}{117}}, \bibinfo{pages}{8743--8749}
  (\bibinfo{year}{2020}).

\bibitem{li2016nitrogen}
\bibinfo{author}{Li, Y.-F.}, \bibinfo{author}{Marty, B.},
  \bibinfo{author}{Shcheka, S.}, \bibinfo{author}{Zimmermann, L.} \&
  \bibinfo{author}{Keppler, H.}
\newblock \bibinfo{journal}{\bibinfo{title}{Nitrogen isotope fractionation
  during terrestrial core-mantle separation}}.
\newblock {\emph{\JournalTitle{Geochemical Perspectives Letters}}}
  \textbf{\bibinfo{volume}{2}} (\bibinfo{year}{2016}).

\bibitem{speelmanns2019}
\bibinfo{author}{Speelmanns, I.~M.}, \bibinfo{author}{Schmidt, M.~W.} \&
  \bibinfo{author}{Liebske, C.}
\newblock \bibinfo{journal}{\bibinfo{title}{The almost lithophile character of
  nitrogen during core formation}}.
\newblock {\emph{\JournalTitle{Earth and Planetary Science Letters}}}
  \textbf{\bibinfo{volume}{510}}, \bibinfo{pages}{186--197}
  (\bibinfo{year}{2019}).

\bibitem{grewal2019fate}
\bibinfo{author}{Grewal, D.~S.} \emph{et~al.}
\newblock \bibinfo{journal}{\bibinfo{title}{The fate of nitrogen during
  core-mantle separation on {E}arth}}.
\newblock {\emph{\JournalTitle{Geochimica et Cosmochimica Acta}}}
  \textbf{\bibinfo{volume}{251}}, \bibinfo{pages}{87--115}
  (\bibinfo{year}{2019}).

\bibitem{shibazaki2009}
\bibinfo{author}{Shibazaki, Y.}, \bibinfo{author}{Ohtani, E.},
  \bibinfo{author}{Terasaki, H.}, \bibinfo{author}{Suzuki, A.} \&
  \bibinfo{author}{Funakoshi, K.}
\newblock \bibinfo{journal}{\bibinfo{title}{Hydrogen partitioning between iron
  and ringwoodite: {I}mplications for water transport into the {M}artian
  core}}.
\newblock {\emph{\JournalTitle{Earth and Planetary Science Letters}}}
  \textbf{\bibinfo{volume}{287}}, \bibinfo{pages}{463--470}
  (\bibinfo{year}{2009}).

\bibitem{li2020}
\bibinfo{author}{Li, Y.}, \bibinfo{author}{Vo{\v{c}}adlo, L.},
  \bibinfo{author}{Sun, T.} \& \bibinfo{author}{Brodholt, J.~P.}
\newblock \bibinfo{journal}{\bibinfo{title}{The {E}arth’s core as a reservoir
  of water}}.
\newblock {\emph{\JournalTitle{Nature Geoscience}}}
  \textbf{\bibinfo{volume}{13}}, \bibinfo{pages}{453--458}
  (\bibinfo{year}{2020}).

\bibitem{tagawa2021}
\bibinfo{author}{Tagawa, S.} \emph{et~al.}
\newblock \bibinfo{journal}{\bibinfo{title}{Experimental evidence for hydrogen
  incorporation into {E}arth’s core}}.
\newblock {\emph{\JournalTitle{Nature communications}}}
  \textbf{\bibinfo{volume}{12}}, \bibinfo{pages}{1--8} (\bibinfo{year}{2021}).

\bibitem{wetzel2013}
\bibinfo{author}{Wetzel, D.~T.}, \bibinfo{author}{Rutherford, M.~J.},
  \bibinfo{author}{Jacobsen, S.~D.}, \bibinfo{author}{Hauri, E.~H.} \&
  \bibinfo{author}{Saal, A.~E.}
\newblock \bibinfo{journal}{\bibinfo{title}{Degassing of reduced carbon from
  planetary basalts}}.
\newblock {\emph{\JournalTitle{Proceedings of the National Academy of
  Sciences}}} \textbf{\bibinfo{volume}{110}}, \bibinfo{pages}{8010--8013}
  (\bibinfo{year}{2013}).

\bibitem{stanley2014}
\bibinfo{author}{Stanley, B.~D.}, \bibinfo{author}{Hirschmann, M.~M.} \&
  \bibinfo{author}{Withers, A.~C.}
\newblock \bibinfo{journal}{\bibinfo{title}{Solubility of {C--O--H} volatiles
  in graphite-saturated martian basalts}}.
\newblock {\emph{\JournalTitle{Geochimica et Cosmochimica Acta}}}
  \textbf{\bibinfo{volume}{129}}, \bibinfo{pages}{54--76}
  (\bibinfo{year}{2014}).

\bibitem{libourel2003}
\bibinfo{author}{Libourel, G.}, \bibinfo{author}{Marty, B.} \&
  \bibinfo{author}{Humbert, F.}
\newblock \bibinfo{journal}{\bibinfo{title}{Nitrogen solubility in basaltic
  melt. {P}art {I}. {E}ffect of oxygen fugacity}}.
\newblock {\emph{\JournalTitle{Geochimica et Cosmochimica Acta}}}
  \textbf{\bibinfo{volume}{67}}, \bibinfo{pages}{4123--4135}
  (\bibinfo{year}{2003}).

\bibitem{boulliung2020}
\bibinfo{author}{Boulliung, J.} \emph{et~al.}
\newblock \bibinfo{journal}{\bibinfo{title}{Oxygen fugacity and melt
  composition controls on nitrogen solubility in silicate melts}}.
\newblock {\emph{\JournalTitle{Geochimica et Cosmochimica Acta}}}
  \textbf{\bibinfo{volume}{284}}, \bibinfo{pages}{120--133}
  (\bibinfo{year}{2020}).

\bibitem{abe1988}
\bibinfo{author}{Abe, Y.} \& \bibinfo{author}{Matsui, T.}
\newblock \bibinfo{journal}{\bibinfo{title}{Evolution of an impact-generated
  {H}$_2${O}--{C}{O}$_2$ atmosphere and formation of a hot proto-ocean on
  {E}arth}}.
\newblock {\emph{\JournalTitle{Journal of the Atmospheric Sciences}}}
  \textbf{\bibinfo{volume}{45}}, \bibinfo{pages}{3081--3101}
  (\bibinfo{year}{1988}).

\bibitem{kasting1988}
\bibinfo{author}{Kasting, J.~F.}
\newblock \bibinfo{journal}{\bibinfo{title}{Runaway and moist greenhouse
  atmospheres and the evolution of {E}arth and {V}enus}}.
\newblock {\emph{\JournalTitle{Icarus}}} \textbf{\bibinfo{volume}{74}},
  \bibinfo{pages}{472--494} (\bibinfo{year}{1988}).

\bibitem{nakajima1992}
\bibinfo{author}{Nakajima, S.}, \bibinfo{author}{Hayashi, Y.-Y.} \&
  \bibinfo{author}{Abe, Y.}
\newblock \bibinfo{journal}{\bibinfo{title}{A study on the “runaway
  greenhouse effect” with a one-dimensional radiative--convective equilibrium
  model}}.
\newblock {\emph{\JournalTitle{Journal of Atmospheric Sciences}}}
  \textbf{\bibinfo{volume}{49}}, \bibinfo{pages}{2256--2266}
  (\bibinfo{year}{1992}).

\bibitem{kurokawa2018subduction}
\bibinfo{author}{Kurokawa, H.}, \bibinfo{author}{Foriel, J.},
  \bibinfo{author}{Laneuville, M.}, \bibinfo{author}{Houser, C.} \&
  \bibinfo{author}{Usui, T.}
\newblock \bibinfo{journal}{\bibinfo{title}{Subduction and atmospheric escape
  of {E}arth's seawater constrained by hydrogen isotopes}}.
\newblock {\emph{\JournalTitle{Earth and Planetary Science Letters}}}
  \textbf{\bibinfo{volume}{497}}, \bibinfo{pages}{149--160}
  (\bibinfo{year}{2018}).

\bibitem{walker1981}
\bibinfo{author}{Walker, J.~C.}, \bibinfo{author}{Hays, P.} \&
  \bibinfo{author}{Kasting, J.~F.}
\newblock \bibinfo{journal}{\bibinfo{title}{A negative feedback mechanism for
  the long-term stabilization of {E}arth's surface temperature}}.
\newblock {\emph{\JournalTitle{Journal of Geophysical Research: Oceans}}}
  \textbf{\bibinfo{volume}{86}}, \bibinfo{pages}{9776--9782}
  (\bibinfo{year}{1981}).

\bibitem{kasting1993}
\bibinfo{author}{Kasting, J.~F.}
\newblock \bibinfo{journal}{\bibinfo{title}{Earth's early atmosphere}}.
\newblock {\emph{\JournalTitle{Science}}} \bibinfo{pages}{920--926}
  (\bibinfo{year}{1993}).

\bibitem{murphykoop2005}
\bibinfo{author}{Murphy, D.~M.} \& \bibinfo{author}{Koop, T.}
\newblock \bibinfo{journal}{\bibinfo{title}{Review of the vapour pressures of
  ice and supercooled water for atmospheric applications}}.
\newblock {\emph{\JournalTitle{Quarterly Journal of the Royal Meteorological
  Society}}} \textbf{\bibinfo{volume}{131}}, \bibinfo{pages}{1539--1565}
  (\bibinfo{year}{2005}).

\bibitem{stolper1988}
\bibinfo{author}{Stolper, E.} \& \bibinfo{author}{Holloway, J.~R.}
\newblock \bibinfo{journal}{\bibinfo{title}{Experimental determination of the
  solubility of carbon dioxide in molten basalt at low pressure}}.
\newblock {\emph{\JournalTitle{Earth and Planetary Science Letters}}}
  \textbf{\bibinfo{volume}{87}}, \bibinfo{pages}{397--408}
  (\bibinfo{year}{1988}).

\end{thebibliography}

\section*{Acknowledgements (not compulsory)}
This work was supported by JSPS KAKENHI Grant No. 19J21445 and 18K13602, and MEXT KAKENHI Grant No. JP17H06457.

\section*{Author contributions statement}
H.S., H.K., H.G., and K.O. designed the project\pr{;} H.S. performed the numerical simulations and analysed the results\pr{;} H.S. and H.K. interpreted the results and wrote the manuscript.

\section*{Additional information}
\textbf{Competing interests}: The authors declare no competing interests.
\section*{Code availability}
The codes used to generate these results and \pr{the} data that support the findings of this study are available at\\
http://www.geo.titech.ac.jp/lab/okuzumi/sakuraba/Contents.html.

\end{document}